\documentclass[12pt,twoside]{article}
\usepackage{amsmath}
\usepackage{amsfonts,amssymb,latexsym}
\usepackage{amscd}
\usepackage{eucal,mathrsfs}
\usepackage{amsthm}
\usepackage{graphics}
\def\Bbb{\mathbb}
\newtheorem{theorem}{{\rm THEOREM}}[section] 
\newtheorem{corollary}[theorem]{{\rm COROLLARY}} 

\newtheorem{lemma}[theorem]{{\rm LEMMA}} 
\newtheorem{proposition}[theorem]{{\rm PROPOSITION}} 
 
\newtheorem{definition}[theorem]{{\rm DEFINITION}}%[section]
\theoremstyle{remark}
\newtheorem{example}{Example}[section]
\newtheorem{remark}{Remark}[section]
\newtheorem{remarks}{Remarks}[section]
\newtheorem{note}{Comment}[section]
\numberwithin{equation}{section}
\newcommand{\cqd}{\hfill \blacktriangle}
\newcounter{contador}

\newcommand{\es}{\\[3mm]}
\def\un{1\kern-3pt \rm I}
\def\ptoday{{\ifcase\month 
\or January, \or February, \or March, \or April,\or May, 
\or June, \or July, \or August, \or September, \or October, 
\or November, \or December,\fi\ \number \year}}
\begin{document}

\title{Supersymmetric Field-Theoretic Models \\ on a Supermanifold}

\author{\normalsize{\bf Daniel H.T. Franco}\footnote{Present address: Universidade Federal
do Esp\'\i rito Santo (UFES), Departamento de F\'\i sica, Campus Universit\'ario de
Goiabeiras, Vit\'oria, ES, Brasil.}\,\,\footnote{e-mail:
dhtf@terra.com.br}\,\,$\textsuperscript{$(a)$}$ and 
{\bf Caio M.M. Polito}$\textsuperscript{$(b)$}$ \\
\\
{\normalsize {$(a)$} {\em Centro de Estudos de F\'\i sica Te\'orica}} \\
{\normalsize {\em Rua Rio Grande do Norte 1053/302 - Funcion\'arios}}\\
{\normalsize {\em CEP:30130-131 - Belo Horizonte - MG - Brasil}.}\\
\\
{\normalsize {$(b)$} {\em Centro Brasileiro de Pesquisas F\'{\i}sicas -- (CBPF)}} \\ 
{\normalsize {\em Coordena\c c\~ao de Campos e Part\'{\i}culas -- (CCP)}} \\
{\normalsize {\em Rua Dr. Xavier Sigaud 150 - Urca}}\\
{\normalsize {\em CEP:22290-180 - Rio de Janeiro - RJ - Brasil}.}} 

\date{}

\maketitle

\vspace{-1cm}

\begin{abstract}
We propose the extension of some structural aspects that have successfully been
applied in the development of the theory of quantum fields propagating on a general
spacetime manifold so as to include superfield models on a supermanifold. We
only deal with the limited class of supermanifolds which admit the existence of a
smooth body manifold structure. Our considerations are based on the
Catenacci-Reina-Teofillatto-Bryant approach to supermanifolds. In particular, we
show that the class of supermanifolds constructed by Bonora-Pasti-Tonin satisfies
the criterions which guarantee that a supermanifold admits a Hausdorff body manifold.
This construction is the closest to the physicist's intuitive view of superspace as
a manifold with some anticommuting coordinates, where the odd sector is topologically
trivial. The paper also contains a new construction of superdistributions and useful
results on the wavefront set of such objects. Moreover, a generalization of the spectral
condition is formulated using the notion of the wavefront set of superdistributions,
which is equivalent to the requirement that all of the component fields satisfy, on the
body manifold, a microlocal spectral condition proposed by Brunetti-Fredenhagen-K\"ohler.
\end{abstract}

%\subjclass{35A18; 35A20; 46S60}


%%%%%%%%%%%%%%%%%%%%%%%%%%%%%%%%%%%%%%%%%%%%%%%%%%%%%%%
\section{Introduction}
\hspace*{\parindent}
%%%%%%%%%%%%%%%%%%%%%%%%%%%%%%%%%%%%%%%%%%%%%%%%%%%%%%%
There are topics in the physical literature which do not exhaust themselves, but always
deserve new analyses. Amongst these, the program to a quantum gravity theory has a
significant part, remaining an open problem of Physics and an active area of current
research. In spite of the fact that many attempts have been made to include gravity in
the quantization program, a satisfactory and definitive theory still does not exist.
Many lines of research in quantum gravity developed over last decades, under different
names, such as the Supergravity, Kaluza-Klein, String, Twistors, D-brane, Loop Quantum Gravity,
Noncommutative Geometry and Topos theories, have elucidated the role of quantum gravity,
without, however, providing conclusive results (see for instance~\cite{Carlip} for a recent
review of the status of quantum gravity). Whereas these good ideas stay only as good promises
in the direction of a final theory of the quantum gravity, and since the relevant scale of
the Standard Model, or any of its supersymmetric extensions, is much below the typical
gravity scale, it seems appropriate to treat, in an intermediate step, some aspects
of gravity in quantum field theory by considering the approach which describes the matter
quantum fields under the influence of a gravitational background. This framework has a wide
range of physical applicability, the most prominent being the gravitational effect of particle
creation in the vicinity of black-holes, raised up for the first time by Hawking~\cite{Haw}. 

The study of quantum field theories on a general manifold has become an
area of intensive research activity, and a substantial progress has been
made on a variety of interesting problems. In particular,
great strides have been made towards the understanding of the question of
how the spectral condition can be defined. While the most of the Wightman 
axioms can be implemented on a curved spacetime, the spectral 
condition (which expresses the positivity of the energy) represents a 
serious conceptual problem. On a flat spacetime the Poincar\'e covariance, 
in particular the translations, guarantees the positivity of the spectrum, 
and fixes a unique vacuum state; but on a general curved spacetime,
due the absence of a global Poincar\'e group, there does not exist a useful notion
of a vacuum state. As a result, the concept of particles becomes ambiguous, and the
problem of the physical interpretation becomes much more difficult. One possible
resolution to this difficulty is to choose some quantities other than particles
content to label quantum states. Such an advice was given by 
Wald~\cite{Wald} with the purpose of finding the expectation value of the 
energy-momentum tensor. For free fields, this approach leds to the concept of 
Hadamard states. The latter are thought to be good candidates for describing 
physical states, at least for free quantum field theories in curved 
spacetime, according to the work of DeWitt and Brehme~\cite{WiBr} 
(see~\cite{Ful, KaWa, Wal} for a general review and references). 
In a seminal work, Radzikowski~\cite{Rad} showed that the global 
Hadamard condition can be locally characterized in terms of the wavefront 
set, and proved a conjecture by Kay~\cite{Kay} that a locally Hadamard quasi-free
Klein-Gordon state on any globally hyperbolic curved spacetime must be globally
Hadamard. His proof relies on a general {\em wavefront set 
spectrum condition} for the two-point distribution, which 
has made the connection with the spectral condition much more transparent 
(see also~\cite{Ko,BFK}). 

The wavefront set was introduced by the mathematicians H\"ormander
and Duistermaat around 1971~\cite{Hor1,DH} in their studies on the propagation 
of singularities of pseudodifferential operators, which rely on what is now known 
as a microlocal point of view. This subject is growing of importance, with a
range of
applications going beyond the original problems of linear partial equations. In particular, 
the link with quantum field theories on a curved spacetime is now firmly 
established, specially after Radzikowski's work. A considerable amount of 
recent papers devoted to this subject~\cite{Ko,BFK,Jun}-\cite{SaVe} emphazises
the importance of the microlocal technique to solving some previously unsolved 
problems. 

At the same time, it seems that not so much attention has been drawn to supersymmetric
theories in this direction. Much of the progress made in understanding the physics of
elementary particles has been achieved through a study of supersymmetry. The latter is a
subject of considerable interest amongst physicists and mathematicians. It is not only it
fascinanting in its own right; in the 30 years that have passed since its proposal,
supersymmetry has been studied intensively in the belief that such theories may play a part
in a unified theory of the fundamental forces, and many issues are understood much better now.
Although no clear signal has been observed up to now, supersymmetry is believed to be
detectable, at least if certain minimal models of particle physics turn out to be realized
in nature, and calculations and phenomenological analysis of supersymmetry models are
well-justified in view of the forthcoming generation of machines, as the new super collider
LHC being buit at CERN, which is expected to operate in a few years time and will have probably
enough high energy to reveal some of the predicted supersymmetry particles, such as neutralinos,
sleptons and may be indirectly squarks. It also has proven to be a tool to link the quantum
field theory and noncommutative geometry~\cite{Wi,Jaf}. Furthermore, in recent years the
supersymmetry have been instrumental in uncovering non-perturbative aspects of quantum
theories~\cite{SeWi,Se}. All of this gives strong motivations for trying to get a deeper 
understanding of the structure and of the properties of supersymmetric field theories. 

This work is inspired in the structurally significant, recent results on quantum
fields propagating in a globally hyperbolic, curved spacetime, and represents a 
natural attempting to construct a generalization of some of the conventional
mathematical structures used in quantum field theory, such as manifolds, so as to
include superfield models in supermanifolds (curved superspaces). These structural questions
are not without physical interest and relevance! It is the purpose of the present 
paper to study how such a construction can be achieved.

The outline of the paper is as follows. We shall begin in Sec. 2 by describing some
global properties of supermanifolds according to Rogers~\cite{Rogers}, and the problem of
constructing their bo\-dies in the sense of Catenacci {\em et al.}~\cite{CaReTe} and
Bryant~\cite{Bry}. Then, by working with a class of $G^\infty$ supermanifolds constructed
by Bonora-Pastin-Tonin~\cite{BoPaTo} (BPT-supermanifolds), we demonstrate that this
class of supermanifolds satisfies the criterions which guarantee that a supermanifold
admits a Hausdorff body manifold. In Sec. 3, superdistributions on superspace are defined.
We derive some results not contained in~\cite{NK}. In particular, we generalize
straightforwardly the notion of distributions defined on a manifold to distributions defined
on a supermanifold. In Sec. 4, we discuss the algebraic formalism so as to
include supersymmetry on a supermanifold. The results from this section may be seen as a natural
extension of the ``Haag-Kastler-Dimock'' axioms~\cite{HaKa,Dimo} for local ``ob\-ser\-va\-bles''
to supermanifolds. In Sec. 5, we summarize some basics on the description of Hadamard
(super)states. The focus of the Sec. 6 will be on the extension of the H\"ormander's description
of the singularity structure (wavefront set) of a distribution to include the supersymmetric
case. This fills a gap in the literature between the usual textbook presentation of the
singularity structure of superfunctions and the rigorous mathematical treatment based on
microlocal analysis. In Sec. 7, we present the characterization of a type of microlocal
spectral condition for a superstate $\omega^{\rm susy}$ with $m$-point superdistribution
$\omega_m^{\rm susy}$ on a supermanifold, in terms of the wavefront set of superdistributions,
which is equivalent to the requirement that all of the component fields satisfy the
microlocal spectral conditions~\cite{BFK} on the body manifold. This is in accordance with the
DeWitt's remark~\cite{DeWitt} which asserts that in physical applications of supersymmetric
quantum field theories, the spectral condition of the GNS-Hilbert superspace is restricted to
the ordinary GNS-Hilbert space that sits inside the GNS-Hilbert superspace. Finally, the
Sec. 8 contains ours final considerations.

%%%%%%%%%%%%%%%%%%%%%%%%%%%%%%%%%%%%%%%%%%%%%%%%%%%%%%%
\section{Notions of Supermanifolds}
\label{NoSmf}
\hspace*{\parindent}
%%%%%%%%%%%%%%%%%%%%%%%%%%%%%%%%%%%%%%%%%%%%%%%%%%%%%%%
This section introduces some few basic fundamentals on the theory of
supermanifolds. We follow here the work of Rogers~\cite{Rogers}
which is both general and mathematically rigorous. 
Rogers' theory has an advantage, a supermanifold is an
ordinary Banach manifold endowed with a Grassmann algebra structure, 
so that the topological constructions have their standard meanings. In this
context see also the Refs.~\cite{DeWitt}-\cite{Bruzzo}.

We start by introducing first some definitions and 
concepts of a Grassmann-Banach algebra, i.e., a Grassmann algebra
endowed with a Banach algebra structure. This leads to the key
concept of supercommutative superalgebra.

\begin{definition}
An algebra is said to be a supercommutative superalgebra $\Lambda$
-- or a ${\Bbb Z}_2$-graded commutative algebra -- if $\Lambda$ is
the direct sum $\Lambda=\Lambda_0 \oplus \Lambda_1$
of two complementary subspaces such that ${\un}\in \Lambda_0$ and 
$\Lambda_0\Lambda_0 \subset \Lambda_0$,
$\Lambda_0\Lambda_1 \subset \Lambda_1$,
$\Lambda_1\Lambda_1 \subset \Lambda_0$. 
Moreover, for all homegeneous element $x$, $y$ in $\Lambda$, 
$xy=(-1)^{|x||y|}yx$, where $|x|=0$ if $x \in \Lambda_0$ 
and $|x|=1$ if $x \in \Lambda_1$. In particular, it follows 
that the square of odd elements is zero.
\label{spal}
\end{definition}

Elements from $\Lambda_0$ and $\Lambda_1$ are said to be homegeneous if they
have a definite parity, i.e., an element $x \in \Lambda_0$ 
is said to have {\em even} parity, while an element $x \in \Lambda_1$
is said to have {\em odd} parity. Products of homogeneous elements
of the same parity are even and of elements of different
parities are odd. 

We shall assume that the superalgebra $\Lambda$ is a Banach space
with norm $\|\cdot\|$ satisfying the condition
\[
\|xy\|\leq \|x\| \|y\| \,\,,\forall x,y \in \Lambda;\quad
\|{\un}\|=1\,\,.
\]

Let $L$ be a finite positive integer and ${\mathscr G}$ 
denote a Grassmann algebra, such that ${\mathscr G}$ can naturally be decomposed
as the direct sum ${\mathscr G}={\mathscr G}_0 \oplus {\mathscr G}_1$, where
${\mathscr G}_0$ consists of the even (commuting) elements and ${\mathscr G}_1$
consists of the odd (anti-commuting) elements in ${\mathscr G}$, respectively.
Let $M_L$ denote the set of sequences 
$\{(\mu_1,\ldots,\mu_k)\mid
1 \leq k \leq L; \mu_i \in {\Bbb N}; 1 \leq \mu_1 < \cdots < \mu_k \leq L\}$.
Let $\Omega$ represent the empty sequence in $M_L$, and $(j)$ denote the sequence
with just one element $j$. A basis of ${\mathscr G}$ is given by  monomials of the form 
$\{\xi_\Omega,\xi^{\mu_1}\xi^{\mu_2},\ldots,\xi^{\mu_1} 
\xi^{\mu_2}\ldots \xi^{\mu_k}\}$ for all $\mu \in M_L$, such that 
$\xi_{\Omega}={\un}$ and 
$\xi^{(i)}\xi^{(j)}+\xi^{(j)}\xi^{(i)}=0$ for 
$1\leq i,j \leq L$. Futhermore, there is no other independent relations 
among the generators. By ${\mathscr G}_L$ we 
denote the Grassmann algebra with $L$ generators, where the even and 
the odd elements, respectively, take their values. $L$ being assumed a finite 
integer (the number of generators $L$ could be possibly infinite),
it means that the sequence terminates at $\xi^{1}\ldots \xi^{L}$ and there are
only $2^L$ distinct basis elements. An arbitrary element $q \in {\mathscr G}_L$
has the form
\begin{equation}
q=q_{\bf b} + \sum_{(\mu_1,\ldots,\mu_k)\in M_L}
q_{\mu_1,\ldots,\mu_k}\xi^{\mu_1}\cdots \xi^{\mu_k}\,\,,
\end{equation}
where $q_{\bf b},q_{{\mu_1\ldots \mu_k}}$ are real numbers. An even or odd element is
specified by $2^{L-1}$ real parameters. The number $q_{\bf b}$ is called the body of
$q$, while the remainder $q-q_{\bf b}$ is the soul of $q$, denoted $s(q)$.
The element $q$ is invertible if, and only if, its body is non-zero.

With reference to supersymmetric field theories, the commuting variable $x$ has
the form
\begin{equation}
x=x_{\bf b}+x_{ij}\xi^i \xi^j+x_{ijkl}\xi^i \xi^j \xi^k \xi^l+\cdots\,\,,
\label{snx}
\end{equation}
where $x_{\bf b},x_{ij},x_{ijkl},\ldots$ are real variables.
Similarly, the anticommuting variables (in the Weyl representation) 
$\theta$ and $\bar{\theta}=(\theta)^*$ have the form
\begin{equation}
\theta=\theta_{i}\xi^i+\theta_{ijk}\xi^i \xi^j \xi^k +\cdots\,\,,
\quad
\bar{\theta}=\bar{\theta}_{i}\xi^i+
\bar{\theta}_{ijk}\xi^i \xi^j \xi^k +\cdots\,\,,
\label{snt}
\end{equation}
where $\theta_{i},\theta_{ijk},\ldots$ are complex variables. The summation over repeated
indices is to be understood unless otherwise stated.

\begin{remark}
As pointed out by Vladimirov-Volovich~\cite{Vlavolo}, 
from the physical point of view, superfields are not functions of 
$\theta_{i},\theta_{ijk},\ldots$ and $x_{\bf b},x_{ij},x_{ijkl},\ldots$, 
but only depend on these variables through $\theta$ and $x$,
as it occurs with ordinary complex analysis where analytic functions 
of the complex variables $z=x+iy$ are not arbitrary functions of the 
variables $x$ and $y$, but functions that depend on $x$ and $y$ 
through $z$. $\cqd$
\end{remark}

The Grassmann algebra may be topologized. Consider the complete norm 
on ${\mathscr G}_L$ defined by~\cite{Rudo}: 
\begin{equation}
\|q\|_p=\left(|q_{\bf b}|^{p}+\sum_{(\mu)=1}^L
|q_{{\mu_1 \ldots \mu_k}}|^p\right)^{1/p}\,\,.
\label{norma}
\end{equation}
A useful topology on ${\mathscr G}$ is the 
topology induced by this norm. The norm $\|\cdot\|_1$ is called the Rogers 
norm and ${\mathscr G}_L(1)$ the Rogers algebra~\cite{Rogers}.
The Grassmann algebra ${\mathscr G}$ equipped with the norm (\ref{norma}) 
becomes a Banach space. In fact ${\mathscr G}$ becomes a 
Banach algebra, i.e., $\|{\un}\|=1$ and $\|qq^\prime\| \leq \|q\|\|q^\prime\|$ 
for all $q,q^\prime \in {\mathscr G}$.

\begin{definition}
A Grassmann-Banach algebra is
a Grassmann algebra endowed with a Banach algebra structure.  
\end{definition}

A superspace must be constructed using as a building block a
Grassmann-Banach algebra ${\mathscr G}_L$ and not only a
Grassmann algebra.

\begin{definition}
Let
${\mathscr G}_L={\mathscr G}_{L,0} \oplus {\mathscr G}_{L,1}$ be
a Grassmann-Banach algebra. Then the $(m,n)$-dimensional superspace
is the topological space 
${\mathscr G}_L^{m,n}={\mathscr G}_{L,0}^m \times {\mathscr G}_{L,1}^n$,
which generalizes the space ${\Bbb R}^m$,
consisting of the Cartesian product of $m$ copies of the even 
part of ${\mathscr G}_L$ and $n$ copies of the odd part.
\end{definition}

For an $(m,n)$-dimensional superspace, a typical element of this set
used in physics is denoted by $(z)=(z_1,\ldots,z_{m+n})=
(x_1,\ldots,x_m,\theta_1,\ldots,\theta_{n/2},
\bar{\theta}_1,\ldots,\bar{\theta}_{n/2})$. For instance, for the
$(4,4)$-dimensional Minkowski superspace, which is the space of e.g.
$N=1$ Wess-Zumino model formulated in superfield language and modelled as 
${\mathscr G}_L^{4,4}={\mathscr G}_{L,0}^4 \times {\mathscr G}_{L,1}^4$,
$(z)=(x_1,\ldots,x_4,\theta_1,\theta_{2},\bar{\theta}_1,\bar{\theta}_{2})$. 
The norm on ${\mathscr G}_L^{4,4}$ is defined by 
$\|z\|=\sum_{i=1}^4\|x_i\|+\sum_{j=1}^2\|\theta_j\|+\sum_{k=1}^2\|\bar{\theta}_k\|$.
The topology on ${\mathscr G}_L^{4,4}$ is the topology induced by this norm
-- which is also the product topology.

In supersymmetric quantum field theory, superfields are functions in superspace usually
given by their (terminating) standard expansions in powers of the odd coordinates
\begin{equation}
F(x,\theta,\bar{\theta})=\sum_{(\gamma)=0}^\Gamma f_{(\gamma)}(x)
(\theta)^{(\gamma)}\,\,,
\label{spfield} 
\end{equation}
where $(\theta)^{(\gamma)}$ comprises all monomials in the anticommuting
variables $\theta$ and $\bar{\theta}$ (belonging to odd part of 
a Grassmann-Banach algebra) of degree $|\gamma|$; $f_{(\gamma)}(x)$
is called a component field, whose Lorentz properties are determined by
those of $F(x,\theta,\bar{\theta})$ and by the power $(\gamma)$ of $(\theta)$.
The following notation, extended to more than one $\theta$ variable,
is used (\ref{spfield}): 
$(\theta)=(\theta_1,\bar{\theta}_1,\ldots,\theta_n,\bar{\theta}_n)$, and
$(\gamma)$ is a multi-index $(\gamma_1,\bar{\gamma}_1,\ldots,\gamma_n,\bar{\gamma}_n)$ with
$|\gamma|=\sum_{r=1}^n(\gamma_r+\bar{\gamma}_r)$ and $(\theta)^{(\gamma)}=\prod_{r=1}^n 
\theta_r^{\gamma_r}\bar{\theta}_r^{\bar{\gamma}_r}$. 
In Eq.(\ref{spfield}), for a (4,4)-dimensional superspace, $\Gamma=(2,2)$.

Rogers~\cite{Rogers} considered superfields in ${\mathscr G}_L^{m,n}$ 
as $G^\infty$ superfunctions, i.e., functions whose coefficients $f_{(\gamma)}(x)$ of
their expansions are smooth functions of ${\Bbb R}^m$ into ${\mathscr G}_L$, extended
from ${\Bbb R}^m$ to all of ${\mathscr G}_L^{m,0}$ by $z$-continuation~\cite{Rogers},
which maps functions of real variables into functions of variables in ${\mathscr G}_L^{m,0}$.

\begin{definition}
Let $U$ be an open set in ${\mathscr G}_L^{m,0}$ and let
$\epsilon:{\mathscr G}_L^{m,0} \rightarrow {\Bbb R}^m$ be the body projection
which associates to each $m$-tuple $(x_1,\ldots,x_m) \in {\mathscr G}_L^{m,0}$
an $m$-tuple $(\epsilon(x_1),\ldots,\epsilon(x_m))\in {\Bbb R}^m$.
Let $V$ be an open set in ${\Bbb R}^m$ with $V=\epsilon(U)$.
We get through $z$-continuation -- or ``Grassmann analytic continuation'' --
of a function $f\in C^\infty(V,{\mathscr G}_L)$ a function $z(f)\in G^\infty(U,{\mathscr G}_L)$,
which admits an expansion in powers of the soul of $x$
\begin{align*}
z(f)(x_1,\ldots,x_m)=\sum_{i_1=\cdots=i_m=0}^L \frac{1}{i_1!\cdots i_m!}
\left[\partial_1^{i_1} \cdots \partial_m^{i_m}\right]f(\epsilon(x))
s(x_1)^{i_1} \cdots s(x_m)^{i_m}\,\,,
\end{align*}
where $s(x_i)=(x_i-\epsilon(x_i))$ and $\epsilon(x_i)=(x_i)_{\bf b}$.
\label{galo}
\end{definition}

One should keep always in mind that the continuation involves only the even variables
$z:C^\infty(\epsilon(U))\rightarrow G^\infty(U)$, and that $z(f)(x_1,\ldots,x_m)$ is a
supersmooth function if their components are smooth for soulless values of $x$. 
This justifies the formal manipulations in the physics literature, 
where superfields are manipulated as if their even arguments were ordinary
numbers~\cite{Rabin}: a supersmooth function is completely determined when
its components are known on the body of superspace.

According to Definition \ref{galo}, the superfield $F(x,\theta,\bar{\theta})
\in G^\infty(U,{\mathscr G}_L)$ admits an expansion
\[
F(x,\theta,\bar{\theta})=\sum_{(\gamma)=0}^\Gamma z(f_{(\gamma)})(x)
(\theta)^{(\gamma)}\,\,,
\]
but here with suitable $f_{(\gamma)}\in C^\infty(\epsilon(U),{\mathscr G}_L)$.

Now, we are going to consider some helpful aspects about supermanifolds, based on the
work of Rogers~\cite{Rogers}, replacing the simple superspace ${\mathscr G}_L^{m,n}$ by a
more general supermanifold. Rogers used the concept of $G^\infty$ superfunctions to define
the concept of $G^\infty$ supermanifolds (which can be considered as Banach real manifolds
$C^\infty$ modelled on ${\mathscr G}_L^{m,n}$ of dim $N=2^{L-1}(m+n)$), with a structure
allowing for the definitions of neighbouring points and continuous superfunctions. An
$(m,n)$-dimensional $G^\infty$ supermanifold generalizes the concept of an $m$-dimensional
$C^\infty$ manifold: just as a manifold is a Hausdorff topological space such that every point
has a neighbourhood homeomorphic to ${\Bbb R}^m$ and has local coordinates
$(x_1(p),\ldots,x_m(p))$ in ${\Bbb R}^m$, a supermanifold is a topological space   
which locally looks like ${\mathscr G}_L^{m,n}$ (but not necessarily in its global extent)
and has local coordinates $(x_1(p),\ldots,x_m(p),\theta_1(p),\ldots,\theta_{n}(p))$ in
${\mathscr G}_L^{m,n}$, and whose transition functions fulfill a suitable supersmoothness
condition.

\begin{definition}
A supermanifold is in general a paracompact Hausdorff topological space ${\mathscr M}$,
together with an atlas of charts $\{(X_\alpha,k_\alpha)\mid \alpha \in I\}$, 
over a Grassmann-Banach algebra ${\mathscr G}_L$, 
where the $X_\alpha$ cover ${\mathscr M}$ and each coordinate function $k_\alpha$
is a homeomorphic local maps from $X_\alpha$ onto an open subset 
$\widetilde{X}_\alpha \subset{\mathscr G}_L^{m,n}$, also Hausdorff. 
\end{definition} 

The existence of infinitely differentiable coordinates systems makes
the supermanifold differentiable. The differentiable structure in this topological
space is due to $G^{r}$ ($r=p$ or $p=\infty$) structure of transition functions, 
$k_{\beta}\circ k^{-1}_{\alpha}$, between overlapping coordinate patches,
$k_{\alpha}(X_{\alpha}\cap X_{\beta})$ and $k_{\beta}(X_{\alpha}\cap X_{\beta})$,
required to be supersmooth morphisms for any $\alpha,\beta \in I$. 
The local coordinates are:
\begin{align*}
&u_{i}=p_{i}\circ k_{\alpha}\longmapsto(i=1,\ldots,m)\,\,, \\
&v_{j}=p_{j+m}\circ k_{\alpha}\longmapsto(j=1,\ldots,n)\,\,.
\end{align*} 

In this sense ${\mathscr G}_L^{m,n}$ is an example of $G^{\infty}$ 
supermanifold, unlike of the coarse topology in the DeWitt sense 
\cite{DeWitt} whose structure cannot be even a metric one.  

\begin{definition}  Let $\widetilde{X}_\alpha$ be an open in 
${\mathscr G}_L^{m,n}$ and $f:\widetilde{X}_\alpha \rightarrow {\mathscr G}_L$, 
then: 

\,\,\,(a) $f$ is called $G^{0}$ in $\widetilde{X}_\alpha$ if $f$ is continuous in 
$\widetilde{X}_\alpha$.

\,\,\,(b) $f$ is called $G^{1}$ in $\widetilde{X}_\alpha$ if exist $m+n$ functions 
$G_{k}f:\widetilde{X}_\alpha \rightarrow {\mathscr G}_L$, $k=1,\ldots,m+n$ 
and functions $\eta:{\mathscr G}_L^{m,n} 
\rightarrow {\mathscr G}_L$ such that: 
\begin{align*}
f(a+h,b+k)=&f(a,b)+ \sum_{i=1}^{m}h_{i}\{G_{i}f(a,b)\}
+\sum_{j=1}^{n}k_{j}\{G_{j+m}f(a,b)\}+ \\
&+\parallel h,k \parallel \eta(h,k)\,\,, 
\end{align*}
and $\eta(h,k) \rightarrow 0$ when $\parallel h,k\parallel \rightarrow 0$. 
In this sense, $G_{i}f \rightarrow f^\prime_{i}$. 
\end{definition}

We can generalize to $G^{p}$, with finite $p$ in the following: $f$ is $G^{p}$ in
$\widetilde{X}_\alpha$ if is possible choose $G_{k}f$ which are $G^{p-1}$ with $f \in G^{1}$
em $\widetilde{X}_\alpha$. If it is true to all $p$, $f$ is called $G^{\infty}$. 
In fact, any function which is absolutely convergent (power series) is 
$G^{\infty}$ on $\widetilde{X}_\alpha$, in other words: 
\begin{align*}
&f(z)=\sum_{k_{1}\ldots k_{m+n}=0}^{\infty}a_{k_{1}\ldots k_{m+n}}
z^{k_{1}}_{1}\ldots z^{k_{m+n}}_{m+n}\,\,,\\ 
&f:\widetilde{X}_\alpha \rightarrow {\mathscr G}_{L}, 
\quad \widetilde{X}_\alpha \subset {\mathscr G}_L^{m,n}\quad\mbox{and}\quad 
a_{k_{1}\ldots k_{m+n}} \in {\mathscr G}_L\,\,. 
\end{align*}

Another important fact is the $C^{\infty}$ structure: 
\[
[D^{p}f(z)][\ell^{1},\ell^{2},\ldots,\ell^{p}]= 
\sum_{k_{1}\ldots k_{p}=1}^{m+n}l^{1}_{k_{1}}\ldots l^{p}_{k_{p}}(G_{k_{p}}
G_{k_{p-1}}\ldots G_{k_{1}}f)(z)\,\,,
\]
for all $z \in \widetilde{X}_\alpha$ open in 
${\mathscr G}_L^{m,n}$ and $l^{1}_{k_{1}}\ldots l^{p}_{k_{p}} \in ({\mathscr G}_L^{m,n})^p$. 
The latter denotes a product  space of $p$ copies of ${\mathscr G}_L^{m,n}$. In this
way the $p$ derivative of $f \in {\mathscr L}[({\mathscr G}_L^{m,n})^p,{\mathscr G}_L]$ are
elements of continuous $p$-linear maps of $({\mathscr G}_L^{m,n})^p$ into ${\mathscr G}_L$.
This formalism is interesting and agrees to the H\"{o}rmander's one~\cite{Hor2} (pg.11), where
$f^{(p)} \in L^{p}(X_\alpha,X_\beta)$, are elements of continuous $p$-linear forms from
$X_\alpha$ to $X_\beta$.

\begin{remark}
The discussion of differentiability by Jadczyk-Pilch~\cite{JaPi} is
simpler than the one given by Rogers~\cite{Rogers}. In particular, knowing
already that a function $f$ is a $C^\infty$ map between Banach spaces,
it is needed only to look at its first derivative to know whether $f$ is
supersmooth or not, while according to Rogers an investigation of all derivatives
is necessary. However, the concept of supersmoothness by Jadczyk-Pilch,
and the concept of $G^\infty$ differentiability by Rogers are equivalent. $\cqd$
\end{remark}

%%%%%%%%%%%%%%%%%%%%%%%%%%%%%%%%%%%%%%%%%%%%%%%%%%%%%%%
\subsection{The Body of a Supermanifold}
\hspace*{\parindent}
%%%%%%%%%%%%%%%%%%%%%%%%%%%%%%%%%%%%%%%%%%%%%%%%%%%%%%%
Now that the general idea of structure on a supermanifold has been introduced,
it is time to restrict our attention to the case of fundamental interest: the
problem of constructing the body of a $G^\infty$ supermanifold
which serves as the physical spacetime. Roughly speaking, the body of a 
supermanifold $\mathscr M$ is an ordinary $C^\infty$ spacetime manifold 
${\mathscr M}_0$ obtained from $\mathscr M$ getting rid of all the soul 
coordinates. Because of its extreme generality, Rogers' theory includes many
topologically exotic supermanifolds which are not physically useful, admitting
the possibility of nontrivial topology in the anticommuting directions and 
classes of supermanifolds without a body manifold. But, intuition suggests that only a
bodied $G^\infty$ supermanifold can be physically relevant!

The question of the existence of the body of a supermanifold was clarified
in the papers by R. Catenacci {\em et al.}~\cite{CaReTe} and  P. Bryant~\cite{Bry}. Their
approach is independent of the atlas used, and it is based on the fact that any
$G^\infty$ supermanifold ${\mathscr M}$ admits a foliation ${\mathfrak F}$.
This type of structure is defined and related to the natural notions of
quotient and substructure on a supermanifold. As with many important concepts
in mathematics, there are several equivalent ways of defining the notion
of a foliation. The simplest and most geometric is the following.

\begin{definition}
Let $\mathscr M$ be an $(m,n)$-dimensional supermanifold of class
$G^r$, $0 \leq r \leq p$. A foliation of class $G^r$, and of codimension $m$, is a
decomposition of $\mathscr M$ into disjoint connected subsets
$\{{\mathfrak L}_\alpha\}_{\alpha \in A}$, called the leaves of the foliation,
such that each point of $\mathscr M$ has a neighbourhood $U$ and a system of
$G^r$ coordinates $(x,\theta):U \rightarrow {\mathscr G}_{L,0}^m \times
{\mathscr G}_{L,1}^n$ such that for each leaf ${\mathfrak L}_\alpha$, the components
of $U \cap {\mathfrak L}_\alpha$ are described by surfaces on which all the body
coordinates $\epsilon(x_1),\ldots,\epsilon(x_m)$ are constant. We denote the foliation
by ${\mathfrak F}=\{{\mathfrak L}_\alpha\}_{\alpha \in A}$.
\label{folia} 
\end{definition}

The coordinates referred in the Definition \ref{folia} are said to be distinguished
by the foliation $\mathfrak F$. Under certain regularity conditions on ${\mathfrak F}$,
the quotient space ${\mathscr M}/{\mathfrak F}$ can be given the structure of an ordinary
$m$-dimensional differentiable manifold ${\mathscr M}_0$, which is called the body manifold
of ${\mathscr M}$ (for details see~\cite{CaReTe}). A $G^\infty$ supermanifold whose
${\mathfrak F}$ foliation is regular is called regular itself. On regular
supermanifolds the following theorem holds:

\begin{theorem}[Catenacci-Reina-Teofilatto Theorem]
Let ${\mathscr M}$ be a regular $G^\infty$ supermanifold. Then its body
${\mathscr M}_0$ is a $C^\infty$ manifold. \qed
\label{CRT}
\end{theorem}

As stated by P. Bryant~\cite{Bry}, the necessity of regularity of the soul foliation
in the sense of Catenacci-Reina-Teofilatto is not sufficient to guarantee that a supermanifold
admits a body manifold. He derived necessary and sufficient conditions, namely that leaves
should be closed and do not accumulate, for the existence of a Hausdorff body manifold.

\begin{theorem}[Bryant Theorem 2.5]
Suppose that $\mathscr M$ is a supermanifold. In order that $\mathscr M$ admits a body
manifold, it is necessary and sufficient that the leaves of the soul foliations are closed
in $\mathscr M$ and do not accumulate. \qed
\label{BRY}
\end{theorem}

For our purposes, it will be sufficient to consider the class of $G^\infty$ supermanifolds
constructed by Bonora-Pasti-Tonin~\cite{BoPaTo} (we shall call BPT-supermanifolds for
brevity), which has important applications in theoretical physics -- and fulfills Theorems
\ref{CRT} and \ref{BRY}, as we shall verify presently. These supermanifolds consist of the
Grassmann extensions of {\em any} ordinary $C^\infty$ spacetime manifold. From a given
$m$-dimensional physical spacetime, one constructs first a $(m,0)$-dimensional supermanifold,
and the $(m,n)$-dimensional supermanifold by taking the direct product with
${\mathscr G}_L^{0,n}$. This construction is the closest to the physicist's intuitive view
of superspace as a manifold with some anticommuting coordinates, with the odd Grassmann
variables being topologically trivial.
 
\begin{remark}
As a matter of fact, in any model involving fermions in a general spacetime, the
supermanifold will need to be that constructed from the spinor bundle of the manifold in
the way which we recall now: Let ${\mathscr M}$ be an $m$-dimensional body manifold and
$E$ be an $n$-dimensional vector bundle over ${\mathscr M}$. Suppose that $\{U_\alpha\}$
is a covering of ${\mathscr M}$ by coordinate neighbourhoods which are also trivialisation
neighbouhoods of $E$. Then, the corresponding $(m,n)$-dimensional supermanifold has coordinate
transition functions
\[
x_\alpha^i=\phi_{\alpha\beta}^i(x_\beta)\,\,,
\]
where $\phi_{\alpha\beta}$ is the $z$ continuation of the transition function for $M$ and
\[
\theta_\alpha^i=g_{\alpha\beta\,\,j}^{\quad i}(x_\beta)\theta_\alpha^j\,\,,
\]
with $g_{\alpha\beta}: U_\alpha \cap U_\beta \longrightarrow Gl(n)$ being the transition
function for $E$. It is worthwhile to note that the BPT-supermanifolds are examples
of this construction when the bundle $E$ is trivial. $\cqd$
\end{remark}

For the convenience of the reader, we recall here the construction of
Bonora-Pasti-Tonin~\cite{BoPaTo}. Let $\{(U_\alpha,\psi_\alpha) \mid \alpha \in I\}$
be an atlas for ${\mathscr M}_0$. For each $\alpha \in I$ consider the subset
$X_\alpha$ of the Cartesian product $U_\alpha \times {\mathscr G}_L^{m,0}$ defined by
\begin{align}
{X}_\alpha=\{(x,\bar{x}) \mid x \in U_\alpha, \bar{x} \in {\mathscr G}_L^{m,0},\,\,
{\mbox{and}}\,\, \epsilon(\bar{x})=\psi_\alpha(x)\}\,\,,
\end{align}
and define $k_\alpha:X_\alpha \rightarrow {\mathscr G}_L^{m,0}$ by
$k_\alpha(x,\bar{x})=\bar{x}$ for $(x,\bar{x})\in X_\alpha$. $k_\alpha$
is a homeomorphism and its image is an open subset of ${\mathscr G}_L^{m,0}$.

An important property of the $z$-continuation is the composition of functions.
Let $U$ be an open set in ${\Bbb R}^m$, and let the map $f:{\Bbb R}^m \rightarrow
{\mathscr G}_L^{k,0}$ be represented by the set of $C^\infty$ functions
$\{f_i(x_1,\ldots,x_m), i=1,\ldots,m\}$. Define $z(f)$ as the set of functions $\{z(f_i)\}$.
Let $V$ be an open set in ${\Bbb R}^n$, and consider the maps $f:U \rightarrow V$ and
$g:V^\prime \rightarrow {\mathscr G}_L^{k,0}$, respectively, where $V^\prime \subseteq V$,
and both $f,g$ are $C^\infty$ functions. Then
\begin{equation}
z(g \circ f)=z(g)\circ z(f)\,\,.
\label{compo}
\end{equation} 
Now consider the disjoint union $M=\bigcup_{\alpha \in I}X_\alpha$. Two points of $M$
are equivalent if and only if $(x,\bar{x}) \sim (x^\prime,\bar{x}^\prime)$, such that
$(x,\bar{x})\in X_\alpha$ and $(x^\prime,\bar{x}^\prime)\in X_\beta$ and
$x=x^\prime$, $\bar{x}^\prime=z(\psi_\beta \circ \psi_\alpha^{-1})(\bar{x})$.
Of course $M$ is a Hausdorff space. Then consider the space
${\mathscr M}_G$ equal to the space $M$ modulo the equivalence relation above. The
$k_\alpha$'s provide ${\mathscr M}_G$ with a $G^\infty$ differentiability structure, so that
${\mathscr M}_G$ is a $G^\infty$ $(m,0)$ supermanifold. Let
$\pi_G:{\mathscr M}_G \rightarrow {\mathscr M}_0$ be a continuous and open projection. Locally
$\pi_G \left.\right|_{X_\alpha}(x,\bar{x})=x$ for $(x,\bar{x})\in X_\alpha$.
Since ${\mathscr M}_G$ is a {\em regular} supermanifold, we find straightforwardly that
$\pi_G \circ k_\alpha^{-1}=\psi_\alpha^{-1} \circ \epsilon$ for $\bar{x}\in k_\alpha(X_\alpha)$.
This can be expressed by the commutative diagram: 
\begin{equation*}
\begin{CD}
{X_\alpha} @<{k_\alpha^{-1}}<< {{\mathscr G}_L^{m,0}} \\
@V{\pi_G}VV  @VV{\epsilon}V \\
{U_\alpha} @<{\psi_\alpha^{-1}}<< {\Bbb R}^m
\end{CD}
\end{equation*}

Finally, we construct the $(m,n)$-dimensional supermanifold ${\mathscr M}$ by taking the
direct product of ${\mathscr M}_G$ with ${\mathscr G}_L^{0,n}$. The projection $\pi_S:
{\mathscr M} \rightarrow {\mathscr M}_0$ is the composite map $\pi_G \circ \gamma$, where
$\gamma:{\mathscr M} \rightarrow {\mathscr M}_G$ is the projection onto the first factor.
The map $\gamma$ is $G^\infty$, unlike $\pi_G$ which is a $C^\infty$ function but not a
$G^\infty$.

\begin{corollary} Let $\mathscr M$ be a BPT-supermanifold. Then the leaves of the soul
foliation are regular, closed in ${\mathscr M}$ and do not accumulate. 
\end{corollary}

\begin{proof}
First of all, it is worthwhile noticing that, according to the construction of
Bonora-Pasti-Tonin, two points of a BPT-supermanifold are in the same leaf if, and only if,
they are equivalents in the sense defined above. Then the soul foliation can be defined by
${\mathscr M}/\!\!\sim \overset{\text{def}}{=}{\mathscr M}/{\mathfrak F}$. Once verified the
corollary, we see that a BPT-supermanifold possesses an ordinary body manifold defined by soul
foliation ${\mathscr M}_0 \overset{\text{def}}{=}{\mathscr M}/{\mathfrak F}$,
where ${\mathscr M}_0$ denotes the body manifold. 

In order to show that the leaves of a BPT-supermanifold are closed, the
following considerations are needed: we say that the soul foliation of a BPT-supermanifold is
a Hausdorff space, and that the structure of their supermanifold is regular. This can be
verified through the following theorem by Bryant~\cite{Bry} (Theorem 3.2): Suppose that
$\mathscr M$ is a supermanifold of dimension $(m,n)$ and $\Gamma = \{U_i,\phi_i\}$ is a good
atlas; then the following conditions are equivalent: ({\em i}) $\Gamma=\{U_i,\phi_i\}$
is a regular superstructure on $\mathscr M$, ({\em ii}) when $s$ and $t$ lie in $U_i$,
$s \approx t$ implies $s \sim t$ and ({\em iii}) the body map $\epsilon:{\mathscr M}\rightarrow
{\mathscr M}/{\mathfrak F}$ is locally modelled on $\epsilon_0:B^{m,n} \rightarrow {\Bbb R}^m$
in the sense that exist homeomorphisms $\bar{\phi}_i:\epsilon U_i \rightarrow \epsilon_0
\phi_iU_i$ such that $\bar{\phi}_i \circ \epsilon_{|U_I}=\epsilon_0 \circ \phi$.
When these conditions are satisfied, ${\mathscr M}/{\mathfrak F}$ is Hausdorff and is a smooth
manifold of dimension $m$ with charts $\{\epsilon U_i, {\bar{\phi}}_i\}$. 
For the case of the equivalence relation $(s \sim t)$ of a BPT-supermanifold, we see that it
must be $\approx$ in the Bryant sense because embodies $\sim$ and is transitive. Then $\approx$
implies $\sim$ on the same charts. This means that the conditions of the Theorem 3.2 by Bryant
must be properties of the BPT foliation, and hence is Hausdorff and regular.
Now, the fact that the leaves of a BPT-supermanifold are closed is clear: each point 
($\epsilon(s))$ of ${\mathscr M}/\!\!\sim$ is closed, given that the BPT-supermanifolds
is a Hausdorff space, and the inverse application theorem guarantees that a leaf is
necessarily closed, since being $F$ the leaf in ${\mathscr M}$, $F= \epsilon^{-1}\epsilon(s)$
where $\epsilon^{-1}$ is a continuous map. 

Finally, we shall verify that the leaves of a BPT-supermanifod do not accumulate. First,
we shall suppose that the leaves of soul foliation accumulate~\cite{Folhas} in a
given pair of points, eg $s_+,s_- \in {\mathscr M}$. Note that as ${\mathscr M}/{\mathfrak F}$
is Hausdorff, given two points $x \in {\mathscr M}/{\mathfrak F}$ and
$y \in {\mathscr M}/{\mathfrak F}$ with $x \not= y$, we can separate them by disjoint open
sets. Choice, for example, $\epsilon s_+ = x$ and $\epsilon s_- = y$, where
$\epsilon:{\mathscr M} \rightarrow {\mathscr M}/{\mathfrak F}$. Then, we also can choose
$s_+ \in F^\prime \cup \Sigma_+$ (a transverse submanifold) and
$s_-\in F^\prime \cup \Sigma_-$ (another transverse submanifold). If this is true, 
$s_+,s_-$ must be in the same leaf, by indicating that $\epsilon s_+ =\epsilon s_-$
contradicting the statement which a soul foliation is Hausdorff. Hence, the leaves do not
accumulate. In order to complete the prove, we examine the condition
$\epsilon s_+ = \epsilon s_-$. Due the possibility of choosing arbitrary transverse
submanifolds, we select $\Sigma(s)$ and $\Sigma(t)$ through the some disjoint neighbourhoods
of $s$ and $t$ resp. such that does not exist a $U_i$ which intersects $\Sigma(s)$ and
$\Sigma(t)$. But $\epsilon s_+ =\epsilon s_-$ implies that $s$ and $t$ are in the same chart
$U_i$, so the leaves do not accumulate since $\Sigma(s) \cup \Sigma(t) = \emptyset$.
\end{proof}

The existence of a body manifold places us in a position to consider physically
interpretable field theories on supermanifolds. In order to establish applicability in a
physical system, we need to impose some restrictions regarding to the body manifold 
${\mathscr M}_0$, associated with the supermanifold ${\mathscr M}$. Apart from another aspects,
the causality principle plays a crucial role in our construction. Therefore, we restrict our
body manifold, $({\mathscr M}_0,g_0)$, to be globally hyperbolic Lorentz manifold, by consisting
of a 4-dimensional smooth manifold ${\mathscr M}_0$ (any dimension would be possible) that
can be smoothly foliated by a family of acausal Cauchy surfaces~\cite{KaWa} and a smooth metric
$g_0$ with signature $(+,-,-,-)$. This means that the body manifold must be topologically
equivalent to the Cartesian product of ${\Bbb R}$ and a smooth spacelike hypersurface $\Sigma$
(a Cauchy surface). $\Sigma$ intersects any endless timelike curve at most once. A 4-dimensional
globally hyperbolic Lorentz manifold is orientable and time orientable, i.e., at each
$x \in {\mathscr M}_0$ we may designate a future and past light cone continuously. Moreover,
${\mathscr M}_0$ is assumed to have a spin structure, so that one can consider spinors defined
on it. It can be shown that a 4-dimensional globally hyperbolic Lorentz manifold
admits a spin structure~\cite{Geroch}. In fact, Geroch~\cite{Geroch} pointed out that a
{\em noncompact}, {\em parallelizable} 4-dimensional manifold admits a spin structure. Geroch's
parallelizability criterion applies to a 4-dimensional globally hyperbolic Lorentz manifold.

\begin{remark}
As it has been emphasized in \cite{Ko}, a natural background geometry
that admits a supersymmetric extension of its isometry group can only be of the 
Anti-De-Sitter (AdS) type. In other words, the global supersymmetry should not be compatible
with most spacetimes, an exception being the AdS space. This requirement seems to be an
extremely restrictive condition, since the AdS space has problems with closed
time-like curves, apparently violating causality and leading to problems during 
quantization. Namely, boundary conditions at infinity are needed. Nevertheless, one should
remind that this result refers to extended supergravity theories with gauged $SO(N)$ internal
symmetry~\cite{BrFree}; this is not, however, our case in this paper. Furthermore, this result
can mainly be justified by the heuristic form of introducing the superspace (which may be
bypassed taking into account the Rogers' theory of a global supermanifold). As stressed by
Bruzzo~\cite{Bruzzo}, ``$\ldots$the usual ways of dealing with superspace field theories are
highly unsatisfactory from a mathematical point of view. The superspace is defined formally,
and, for instance, general coordinate transformations are mathematically not well defined.
As a consequence, there is now room for studying global topological properties of superspace.''
As it shall be tackled further on, Section 4, the mathematical structure of the supermanifolds
chosen here leads to a natural formulation of superdiffeormorphisms, $G^\infty$, from
$({\mathscr M},g)$ to $({\mathscr M}^\prime,g^\prime)$, from the $z$-continuation of ordinary
diffeomorphisms, so that these structures become, projectively, well-defined isometries
whenever ${\mathscr M}^\prime={\mathscr M}$ and restricted to the ordinary body manifold. $\cqd$   
\end{remark}

%%%%%%%%%%%%%%%%%%%%%%%%%%%%%%%%%%%%%%%%%%%%%%%%%%%%%%%
\section{Superdistributions}
\label{DiSmf}
\hspace*{\parindent}
%%%%%%%%%%%%%%%%%%%%%%%%%%%%%%%%%%%%%%%%%%%%%%%%%%%%%%%
In this section, as a natural next step, we extend the definition of the
objects most widely used in physics: distributions. We define superdistributions on
supermanifolds over the Grassmann-Banach algebra ${\mathscr G}_L$, as continuous linear
mappings to ${\mathscr G}_L$ from the test function space of $G^\infty$ superfunctions
with compact support. We derive some results not contained in~\cite{NK}.

%%%%%%%%%%%%%%%%%%%%%%%%%%%%%%%%%%%%%%%%%%%%%%%%%%%%%%%

\subsection{Distributions on a Manifold}
\hspace*{\parindent}
%%%%%%%%%%%%%%%%%%%%%%%%%%%%%%%%%%%%%%%%%%%%%%%%%%%%%%%
To prepare for the extension of the theory of distributions to supermanifolds,
we first consider their definition on manifolds. Following~\cite{Hor2}, the spacetime
manifold ${\mathscr M}_0$ (here ${\mathscr M}_0$ denotes an ordinary manifold obtained from
a supermanifold ${\mathscr M}$ by throwing away all the soul coordinates) is a Hausdorff
space covered by charts $(X_\alpha,k_\alpha)$, where the open sets $X_\alpha$ are
homeomorphic neighbourhoods to open sets in ${\Bbb R}^n$. A $C^\infty$ structure on 
${\mathscr M}_0$ is a family ${\mathcal F}=\{(X_\alpha,k_\alpha) \mid \alpha \in I\}$, 
called an atlas, of homeomorphisms $k_\alpha$, called coordinate functions, of open sets 
$X_\alpha \subset {\mathscr M}_0$ on open sets $\widetilde{X}_\alpha \subset {\Bbb R}^n$,
such that $(i)$ if $k_\alpha,k_\beta \in {\mathcal F}$, then the map 
$k_\beta \circ k_\alpha^{-1}:k_\alpha(X_\alpha \cap X_\beta) 
\rightarrow k_\beta(X_\alpha \cap X_\beta)$ is infinitely differentiable, $(ii)$
${\mathscr M}_0=\bigcup_{\alpha \in I} X_\alpha$. Let $f \in C_0^\infty({\Bbb R}^n)$
denotes the set of $C^\infty$ functions of compact support on
$\widetilde{X}_\alpha \subset {\Bbb R}^n$. Then, we can 
represent each $f$ by functions $\bar{f}$ of compact support on 
${\mathscr M}_0$ by $f=\bar{f} \circ k_\alpha^{-1}$, for each $k_\alpha$, 
where $\bar{f} \in C^\infty_0({\mathscr M}_0)$. 
Elements of ${\mathscr D}^\prime({\mathscr M}_0)$, the 
topological dual of $C^\infty_0({\mathscr M}_0)$, are distributions $u$ on 
${\mathscr M}_0$, by which we mean collections 
$\{u_{k_\alpha}\}_{k_\alpha \in {\mathcal F}}$ of 
distributions $u_{k_\alpha} \in {\mathscr D}^\prime(\widetilde{X}_\alpha)$ 
such that $u$ is uniquely 
determined by the $u_{k_\alpha}$ and relations 
$u=u_{k_\alpha} \circ k_\alpha$. Moreover, since for 
any other coordinate system one has $u=u_{k_\beta} \circ k_\beta$ in 
$(X_\alpha \cap X_\beta)$, it follows that $u_{k_\beta}=
(k_\alpha \circ k_\beta^{-1})^{*}u_{k_\alpha}=
u_{k_\alpha}\circ(k_\alpha \circ k_\beta^{-1})$ in $(X_\alpha \cap X_\beta)$. 

%%%%%%%%%%%%%%%%%%%%%%%%%%%%%%%%%%%%%%%%%%%%%%%%%%%%%%%
\subsection{Distributions on the Flat Superspace}
\hspace*{\parindent}
%%%%%%%%%%%%%%%%%%%%%%%%%%%%%%%%%%%%%%%%%%%%%%%%%%%%%%%
With the purpose of defining superdistributions on supermanifolds, 
we must first consider superdistributions on an open set $U \subset {\mathscr G}_L^{m,n}$,
where ${\mathscr G}_L^{m,n}$ denotes the flat superspace. We begin by introducing the concept
of superdistributions as the dual space of supersmooth functions in
${\mathscr G}_L^{m,0}$, with compact support, equipped with an appropriate topology,
called {\em test superfunctions}. This can be done relatively
straightforward in analogy to the notion of distributions as the dual space
to the space $C_0^\infty(U)$ of functions on an open set 
$U \subset {\Bbb R}^m$ which have compact support, since the spaces
${\mathscr G}_L^{m,0}$ and ${\mathscr G}_L^{m,n}$ are regarded as
ordinary vector spaces of $2^{L-1}(m)$ and $2^{L-1}(m+n)$ dimensions,
respectively, over the real numbers. 
 
Let $\Omega \subset {\Bbb R}^m$ be an open set. $\Omega=\epsilon(U)$ regarded
as a subset of ${\mathscr G}_L^{m,0}$, it is identified with the body of some
domain in superspace. Let $C_0^\infty(\Omega,{\mathscr G}_L)$ be the space of 
${\mathscr G}_L$-valued smooth functions with compact support in ${\mathscr G}_L$. 
Every function $f \in C_0^\infty(\Omega,{\mathscr G}_L)$ can be expanded in terms
of the basis elements of ${\mathscr G}_L$ as:
\begin{align}
f(x)=\sum_{(\mu_1,\ldots,\mu_k)\in M_L^0}f_{\mu_1,\ldots,\mu_k}(x)
\xi^{\mu_1}\cdots \xi^{\mu_k}\,\,,
\end{align}
where $M_L^0 {\stackrel{\rm def}{=}}\{(\mu_1,\ldots,\mu_k)\mid
0 \leq k \leq L; \mu_i \in {\Bbb N}; 1 \leq \mu_1 < \cdots < \mu_k \leq L\}$
and $f_{\mu_1,\ldots,\mu_k}(x)$ is in the space $C^\infty_0(\Omega)$ of
real-valued smooth functions on $\Omega$ with compact support. Thus, it follows that
the space $C^\infty_0(\Omega,{\mathscr G}_L)$ is isomorphic to the space 
$C^\infty_0(\Omega) \otimes {\mathscr G}_L$~\cite{NK}. In accordance with the Definition
\ref{galo}, the smooth functions of $C_0^\infty(\Omega,{\mathscr G}_L)$ can be extended
from $\Omega \subset {\Bbb R}^m$ to $U \subset {\mathscr G}_L^{m,0}$ by Taylor expansion.

In order to define superdistributions, we need to give a suitable topological structure
to the space $G_0^\infty(U,{\mathscr G}_L)$ of ${\mathscr G}_L$-valued
superfunctions on an open set $U \subset {\mathscr G}_L^{m,0}$ which have compact support.
According to a proposition by Rogers, every $G^\infty$ superfunction on a
compact set $U \subset {\mathscr G}_L^{m,0}$ can be considered as a real-valued
$C^\infty$ function on $U \subset {\Bbb R}^N$, where $N=2^{L-1}(m)$, regarding 
${\mathscr G}_L^{m,0}$ and ${\mathscr G}_L$ as Banach spaces. In fact, the identification
of ${\mathscr G}_L^{m,0}$ with ${\Bbb R}^{2^{L-1}(m)}$ is possible~\cite{CaReTe}.
We have here an example of functoriality. Indeed, let $X$ and $Y$ denote a $G^\infty$
supermanifold and a Banach manifold $C^\infty$, respectively. Then
with each supermanifold $X$ we associate a Banach manifold $Y$, via a {\em covariant}
functorial relation $\lambda:X \rightarrow Y$, and with each $G^\infty$ map $\phi$ defined
on $X$, a $C^\infty$ map $\lambda(\phi)$ defined on $Y$~\cite{CaReTe}. 

Following, we shall first consider only the subset $C^\infty_K$ of $C^\infty_0(U \subset
{\Bbb R}^N)$ which consists of functions with support in a fixed compact set $K$. Since by
construction $C^\infty_K$ is a Banach space, the functions $C^\infty_K$ have a natural topology
given by the finite family of norms
\begin{equation}
\|\phi\|_{K,m}=\sup_{\stackrel{|p|\leq m}{x \in K}}|D^p \phi(x)|\,\,,
\qquad D^p=\frac{\partial^{|p|}}{\partial x^{p_1}_1 \cdots \partial x^{p_m}_m}\,\,, 
\label{snorm} 
\end{equation}
where $p=(p_1,p_2,\ldots,p_m)$ is a $m$-tuple of non-negative integers, and
$|p|=p_1+p_2+\ldots+p_m$ defines the order of the derivative. Next, let $U$ be considered
as a union of compact sets $K_i$ which form an increasing family $\{K_i\}_{i=1}^\infty$, such
that $K_i$ is contained in the interior of $K_{i+1}$. That such family exist follows from the
Lemma 10.1 of~\cite{Treves1}. Therefore, we think of
$C^\infty_0(U \subset {\Bbb R}^N)$ as $\bigcup_i C^\infty_{K_i}(U \subset {\Bbb R}^N)$. We
take the topology of $C^\infty_0(U \subset {\Bbb R}^N)$ to be given by the strict
inductive limit topology of the sequence $\{C^\infty_{K_i}(U \subset {\Bbb R}^N)\}$. Of 
another way, we may define convergence in $C^\infty_0(U \subset {\Bbb R}^N)$ of a sequence of
functions $\{\phi_k\}$ to mean that for each $k$, one has
${\rm supp}\,\,\phi_k \subset K \subset U \subset {\Bbb R}^N$ such that for a function
$\phi \in C^\infty_0(U \subset {\Bbb R}^N)$ we have
$\|\phi -\phi_k\|_{K,m}\rightarrow 0$ as $k \rightarrow \infty$. This notion of convergence
generates a topology which makes $C^\infty_0(U \subset {\Bbb R}^N)$, certainly, a topological
vector space. 

Now, let $\sf F$ and $\sf E$ be spaces of smooth functions with compact support defined on
$U \subset {\mathscr G}_L^{m,0}$ and $U \subset {\Bbb R}^N$, respectively.
If $\lambda:{\sf E}\rightarrow{\sf F}$ is a {\em contravariant} functor which associates
with each smooth function of compact support in ${\sf E}$, a smooth function of compact
support in ${\sf F}$, then we have a map
\begin{align}
\|\phi\|_{K,m} \longrightarrow \|\lambda(\phi)\|_{K,m}\,\,, 
\label{snorm0} 
\end{align}
providing $G_0^\infty(U,{\mathscr G}_L)$ with a limit topology induced by a
finite family of norms.

We now take a result by Jadczyk-Pilch~\cite{JaPi}, later refined by Hoyos et
al~\cite{HoQuRaUr}, which establishes as a natural domain of definition for supersmooth
functions a set of the form $\epsilon^{-1}(\Omega)$, where $\Omega$ is open in ${\Bbb R}^m$.
Let $\epsilon^{-1}(\Omega)$ be the domain of definition for a superfunction
$f \in G_0^\infty(\epsilon^{-1}(\Omega),{\mathscr G}_L)$, where $\epsilon^{-1}(\Omega)$ is
an open subset in ${\mathscr G}_L^{m,0}$ and $\Omega$ is an open subset in ${\Bbb R}^m$, and
let $\widetilde{\phi} \in C^\infty_0(\Omega,{\mathscr G}_L)$ denotes the restriction of
$\phi$ to $\Omega \subset {\Bbb R}^m \subset {\mathscr G}_L^{m,0}$. Then, it follows that 
$(\partial^{p_1}_1 \cdots \partial^{p_m}_m \phi)^{\widetilde{}}= \partial^{p_1}_1 
\cdots \partial^{p_m}_m{\widetilde{\phi}}$, where the derivatives on the
right-hand side are with respect to $m$ real variables. Now, suppose $\Omega=\bigcup_i
\widetilde{K}_i$ where each $\widetilde{K}_i$ is open and has compact closure in
$\widetilde{K}_{i+1}$. It follows that $C^\infty_0(\Omega,{\mathscr G}_L)=\bigcup_i
C^\infty_{\widetilde{K}_i}(\Omega,{\mathscr G}_L)$. Then, one can give
$C^\infty_0(\Omega,{\mathscr G}_L)$ a limit topology induced by finite family of
norms~\cite{NK}
\begin{equation}
\|\widetilde{\phi}\|_{\widetilde{K},m}=\sup_{\stackrel{|p|\leq m}{x \in \widetilde{K}}}
|D^p \widetilde{\phi}(x)|=\sup_{\stackrel{|p|\leq m}{x \in \widetilde{K}}} 
\left\{\sum_{(\mu_1,\ldots,\mu_k)\in M_L^0}
|D^p \widetilde{\phi}_{\mu_1,\ldots,\mu_k}(x)|\right\}\,\,.
\label{snorma} 
\end{equation}

Finally, a suitable topological structure to the space $G_0^\infty(U,{\mathscr G}_L)$ of
${\mathscr G}_L$-valued superfunctions on an open set $U \subset {\mathscr G}_L^{m,n}$
which have compact support, it is obtained immediately by the natural identification of
${\mathscr G}_L^{m,n}$ with ${\Bbb R}^{2^{L-1}(m+n)}$ and by the obvious extension
of the construction above, which allows us define a limit topology induced to
the space $G_0^\infty(U,{\mathscr G}_L)$ by finite family of norms,
\begin{equation}
\|\lambda(\phi)\|_{K,m+n}=\sup_{\stackrel{|p|\leq m+n}{z \in K}}
|D^p (\lambda(\phi))(z)|\,\,,
\qquad D^p=\frac{\partial^{|q|+|r|}}{\partial x^{q_1}_1 \cdots \partial x^{q_m}_m
\partial\theta^{r_1}_1 \cdots \partial\theta^{r_n}_n}\,\,. 
\label{snorm3} 
\end{equation}
The derivatives $\partial^{|q|}/\partial x^{q_1}_1 \cdots \partial x^{q_m}_m$ commute
while the derivatives $\partial^{|r|}/\partial\theta^{r_1}_1\cdots \partial\theta^{r_n}_n$
anticommute, and $|p|=|q|+|r|=\sum_{i=1}^m q_i+\sum_{j=1}^n r_j$ defines the total order of
the derivative, with $r_j=0,1$.

We are now ready to define a superdistribution in an open subset $U$ of ${\mathscr G}_L^{m,n}$.
The set of all superdistributions in $U$ will be denoted by ${\mathfrak D}^\prime(U)$. A
superdistribution is a continuous linear functional $u:G_0^\infty(U)\rightarrow {\mathscr G}_L$,
where $G_0^\infty(U)$ denotes the test superfunction space of $G^\infty(U)$ superfunctions
with compact support in $K \subset U$. The continuity of $u$ on $G_0^\infty(U)$ is equivalent
to its boundedness on a neighbourhood of zero, i.e., the set of numbers $u(\phi)$ is bounded
for all $\phi \in G_0^\infty(U)$. The last statement translates directly into:

\begin{proposition}
A superdistribution $u$ in $U \in {\mathscr G}_{L}^{m,n}$ is a continuous linear functional
on $G^{\infty}_{0}(U)$ if and only if to every compact set $K \subset U$, there exists a
constant $C$ and $(m+n)$ such that 
\[
\left|u(\phi)\right| \leq C 
\sup_{\stackrel{{|p| \leq m+n}}{z \in K}} 
\left|D^{p}(\phi)(z)\right|\,\,, 
\quad \phi \in G^{\infty}_{0}(K)\,\,.
\]
\label{dspdis} 
\end{proposition}

\begin{proof}
First, it is worth keeping in mind that ${\mathscr G}_L$ can be identified with
${\Bbb R}^{2^{L-1}}$~\cite{CaReTe}. In fact, a number system assuming values in some Grassmann
algebra with $L$ generators is specified by $2^{L-1}$ real parameters. Let $\sf F$ and $\sf E$
be spaces of smooth functions with compact support defined on
$K \subset U \subset {\mathscr G}_L^{m,n}$ and
$K \subset U \subset {\Bbb R}^{2^{L-1}(m+n)}$, respectively.
If we have a functorial relation $\lambda:{\sf F}\rightarrow{\sf E}$ and a linear
functional $\widetilde{u}:{\sf E}\rightarrow {\Bbb R}^{2^{L-1}}$, we can compose $\lambda$
with $\widetilde{u}$ to obtain the pullback of $\widetilde{u}$ by $\lambda$, i.e.,
$u=\lambda^*\widetilde{u}=\widetilde{u}\circ \lambda$, and hence a linear functional
$\lambda^*\widetilde{u}:{\sf F}\rightarrow {\Bbb R}^{2^{L-1}}$. Then, the statement follows if
$\widetilde{u}$ is continuous on $\sf E$. But this clear from the Proposition~21.1
of~\cite{Treves1}, which can be applied {\em verbatim} for a functional $\widetilde{u}$ on
${\sf E}$. 
\end{proof} 

%%%%%%%%%%%%%%%%%%%%%%%%%%%%%%%%%%%%%%%%%%%%%%%%%%%%%%%
\subsection{Distributions on a Supermanifold}
\hspace*{\parindent}
%%%%%%%%%%%%%%%%%%%%%%%%%%%%%%%%%%%%%%%%%%%%%%%%%%%%%%%
Next we will obtain an extension of basic results about superdistributions 
on the flat superspace in the case of general supermanifolds.

\begin{definition}  Let ${\mathscr M}$ a $G^{\infty}$ supermanifold. 
For every coordinate system $p_{i} 
\circ k_{\alpha}$ in ${\mathscr M}$ one has a distribution $u_{k_\alpha} 
\in {\mathfrak D}^\prime(\widetilde{X}_\alpha)$ where $\widetilde{X}_{\alpha}$ 
is an open from ${\mathscr G}_L^{m,n}$ such that 
\begin{eqnarray}
\label{def1}
u_{k_\beta}=\{(p_{i} \circ k_\alpha) \circ (k_\beta^{-1}\circ 
p^{-1}_{i})\}^{*}u_{k_\alpha}\,\,,\quad(i=1,\ldots,m+n)\,\,, 
\end{eqnarray}
in $k_\beta(X_\alpha \cap X_\beta)$, where $p_{i}$ is a projection into 
each copies $(i)$ from ${\mathscr G}^{m,n}$, such that $x_{i}=p_{i} \circ 
k_{\alpha}$ and $y_{j}=p_{j+m} \circ k_{\alpha}$, 
with $(i=1,\ldots,m;j=1,\ldots,n)$. 
We call the system $u_{k_\alpha}$ a distribution $u$ in 
${\mathscr M}$. The set of every distribution in ${\mathscr M}$ 
is denoted by ${\mathfrak D}^\prime({\mathscr M})$. 
\end{definition}

\begin{theorem}  Let $\widetilde{X}_{\alpha},\alpha \in I$, be an arbitrary 
family of open sets in ${\mathscr G}_{L}^{m,n}$, and set 
$\widetilde{X}= \bigcup_{\alpha \in I} \widetilde{X}_{\alpha}$. 
If  $u_{\alpha} \in {\mathfrak D}^\prime(\widetilde{X}_{\alpha})$ and 
$u_{\alpha}=u_{\beta}$ in $(\widetilde{X}_{\alpha} \cap \widetilde{X}_{\beta})$ 
for all $\alpha,\beta \in I$, then there  exist one and only one  
$u \in {\mathfrak D}^\prime(\widetilde{X})$ such that $u_{\alpha}$ 
is the restriction of $u$ to $\widetilde{X}_{\alpha}$ for every $\alpha$. 
\label{juju}
\end{theorem}

To prove this theorem, it is interesting to state the following results:

\begin{lemma}  Let  $\widetilde{X}_{1},\ldots,\widetilde{X}_{k}$ be open sets in 
${\mathscr G}_{L}^{m,n}$ and let $\phi \in 
G^{\infty}_{0}(\bigcup_{1}^{k}\widetilde{X}_{\alpha})$. Then one can find $\phi_{\alpha} \in 
G^{\infty}_{0}(\widetilde{X}_{\alpha}), \alpha=1,\ldots,k$, such that $\phi= \sum_{1}^{k} 
\phi_{\alpha}$ and if $\phi \geq 0$ can take all $\phi_{\alpha} \geq 0$. 
\label{jaja}
\end{lemma}

\begin{proof}
We can choose compact sets $K_{1},\ldots,K_{k}$ with $K_{\alpha} 
\subset \widetilde{X}_{\alpha}$, so that  the supp $\phi \subset \bigcup_{1}^{k}K_{\alpha}$. 
(every point in supp $\phi$ has a compact neighbourhood contained in some 
$\widetilde{X}_{\alpha}$, a finite number of such neighbourhoods can be chosen which cover 
all of supp $\phi$. The union of those which belong to $X_{\alpha}$ is a compact 
set  $K_{\alpha} \subset \widetilde{X}_{\alpha}$. Now, if $\widetilde{X}$ is an open set in
${\mathscr G}_{L}^{m,n}$ and $K$ is a compact subset, then one can find $\phi \in 
G^{\infty}_{0}(\widetilde{X})$ with $0 \leq \phi \leq 1$ so that $\phi=1$ in a 
neighbourhood of $K$. So, we can choose $\psi_{\alpha}\in G^{\infty}_{0}(\widetilde{X}_{\alpha})$ 
with $0 \leq \psi_{\alpha} \leq 1$ and $\psi_{\alpha}=1$ in $K_{\alpha}$,  then the 
functions: 
\[
\phi_{1}=\phi\psi_{1},\phi_{2}=\phi\psi_{2}(1-\psi_{1}),\ldots,\phi_{k}= 
\phi\psi_{k}(1-\psi_{1})\ldots(1-\psi_{k-1})\,\,.
\]
have the required properties since 
\[
\sum_{1}^{k} \phi_{\alpha}-\phi=-\phi 
\prod_{1}^{k} (1-\psi_{\alpha})=0\,\,,
\]
because either $\phi$ or some $1-\psi_{\alpha}$ is zero at any point.
\end{proof}

\begin{corollary}  Let $\widetilde{X}_{1},\ldots,\widetilde{X}_{k}$ be open sets in 
${\mathscr G}_{L}^{m,n}$ and $K$ a compact subset $\subset \widetilde{X}_{\alpha}$. 
Then one can find $\phi_{\alpha}\in G^{\infty}_{0}(\widetilde{X}_{\alpha})$ 
so that $\phi_{\alpha} \geq 0$ and $\sum_{1}^{k} \phi_{\alpha} \leq 1$ with equality
in a neighbourhood of $K$. \qed
\label{jojo}
\end{corollary} 

\begin{proof}[Proof of the Theorem \ref{juju}] If $u$ is a distribution, 
then:
\[ 
u(\phi)=\sum u_{\alpha}(\phi_{\alpha})\,\,,\quad \mbox{if}\quad \phi= \sum 
\phi_{\alpha}\quad (\mbox{where}\,\,\,\phi_{\alpha} \in
G^{\infty}_{0}(\widetilde{X}_{\alpha}))\,\,,
\]
and the sum is finite. By the Lemma \ref{jaja}, every $\phi \in 
G^{\infty}_{0}(\widetilde{X})$ can be written as such a sum. If $\sum \phi_{\alpha}=0 
\Rightarrow \sum u_{\alpha}(\phi_{\alpha})=0$, then we conclude that $\sum 
u_{\alpha}(\phi_{\alpha})$ is independent of how we choose the sum. 
Let $K= \bigcup {\rm supp}\,\phi$ compact set $K \subset \widetilde{X}$ and using the 
corollary \ref{jojo}, we can choose $\psi_{\beta} \in G^{\infty}_{0}(\widetilde{X}_{\beta})$ 
such that $\sum \psi_{\beta}=1$ in $K$ and the sum is finite. 
Then  $\psi_{\beta}\phi_{\alpha} \in G^{\infty}_{0}(\widetilde{X}_{\alpha} \cap 
\widetilde{X}_{\beta})$ so $u_{\alpha}(\psi_{\beta}\phi_{\alpha})=
u_{\beta}(\psi_{\beta}\phi_{\alpha})$. Hence
\[
\sum 
u_{\alpha}(\phi_{\alpha})=\sum \sum u_{\alpha}(\phi_{\alpha}\psi_{\beta})= 
\sum \sum u_{\beta}(\phi_{\alpha}\psi_{\beta})=
\sum u_{\beta}(\psi_{\beta} \sum \phi_{\alpha})=0\,\,.
\]
We have showed that if $\sum \phi_{\alpha}=0 \Rightarrow \sum u_{\alpha}(\phi_{\alpha})$ 
is zero, then $u$ is unique. In order to show that $u$ is distribution, 
choose a compact set $K \subset \widetilde{X}$ and a function $\psi_{\beta} 
\in G^{\infty}_{0}(\widetilde{X}_{\beta})$ with $\sum \psi_{\beta}=1$ in $K$ and finite sum.
If $\phi \in G^{\infty}_{0}(K) $ we have $\phi=\sum \phi\psi_{\beta}$ with $\phi\psi_{\beta}
\in G^{\infty}_{0}(\widetilde{X}_{\beta})$ so that the first equation this proof gives 
\[
u(\phi)=\sum u_{\beta}(\phi\psi_{\beta})\,,
\]
but, if $u_{\beta}$ is a distribution, then: 
\[
\left|u_{\beta} (\phi\psi_{\beta})\right|  \leq C
\sup_{\stackrel{{|p| \leq m+n}}{z \in K}} 
\left|D^{p}(\phi\psi_{\beta})(z)\right|\,\,,\quad \phi\psi_{\beta} 
\in G^{\infty}_{0}(\widetilde{X}_{\beta})\,\,,
\]
where sup $D^{p}\phi$ can be estimated in terms of $\phi$, and 
so we conclude that 
\[
\left|u(\phi)\right| \leq C 
\sup_{\stackrel{{|p| \leq m+n}}{z \in K}} 
\left|D^{p}\phi(z)\right|\,\,,\quad 
\phi \in G^{\infty}_{0}(K)\,\,.
\]
This completes our proof.
\end{proof}
       
\begin{theorem}  Let ${\mathcal F}$ an atlas for ${\mathscr M}$.
If for every $p_{i} \circ k \in {\mathcal F}$ one has a distribution $u_{k} \in 
{\mathfrak D}^\prime(\widetilde{X}_{k})$ and the above definition is true when 
$p_{i}\circ k$ and $p^\prime_{i} \circ k^\prime$ belongs to ${\mathcal F}$, then 
there is one, and only one, distribution $u \in {\mathfrak D}^\prime({\mathscr M})$ 
such that $u \circ (k^{-1}\circ p^{-1}_{i})=u_{k}$ for every $p_{i} \circ k 
\in {\mathcal F}$. 
\end{theorem}

\begin{proof}
Let $\psi \in G^{\infty}$ be a coordinate system in 
${\mathscr M}$. The Theorem \ref{juju} states that there exists one, and only one, 
distribution $U_{\psi} \in {\mathfrak D}^\prime(\widetilde{X}_{\psi})$ in such a
way for every $p_{i} \circ k$, $U_{\psi}= ((p_{i} \circ k) \circ 
\psi^{-1})^{*}u_{k}$ in $\psi(X_{\psi} \cap X_{k}) \subset \widetilde{X}_{\psi}$.
If $\psi \in {\mathcal F} \rightarrow U_{\psi}=u_{\psi}$, we can choose $p_{i}\circ k=\psi$.
Now, one defines $u$ as a distribution, since $U_{\psi}$ satisfies (\ref{def1}) for both
coordinate systems $p_{i} \circ k$ and $p^\prime_{i} \circ k^\prime$.
\end{proof}

%%%%%%%%%%%%%%%%%%%%%%%%%%%%%%%%%%%%%%%%%%%%%%%%%%%%%%%
\section{Algebraic Framework on a Supermanifold}
\hspace*{\parindent}
%%%%%%%%%%%%%%%%%%%%%%%%%%%%%%%%%%%%%%%%%%%%%%%%%%%%%%%
In the usual treatment of quantum field theory in flat spacetime,
the existence of a unitary representation of the restricted Poincar\'e group,
${\mathscr P}_+^\uparrow$, with generators $P_\mu$ fulfilling the spectral condition 
${\sf sp}P_\mu \subset \overline{V}_+$, is very essential. This unitary operator plays a
key role in picking out a preferred vacuum state, i.e., a state which is invariant under all
translations. We choose a complete system of physical states, with positive energies, just
when it is possible to define this vacuum state and consequently the Fock Space, ${\mathscr F}$.
One then defines observables as operators on ${\mathscr F}$ which act upon the states.
However, the characterization of the vacuum involves global aspects, and in the case
of a curved spacetime it is not evident how to select a distinguished state. As already
mentioned in the Introduction, due the absence of a {\em global} Poincar\'e group
there is no analogous selection criterium on a curved spacetime: no vacuum state can be used
as reference. To understand the significance of this point under another point of view, we
take into account that, initially, a theory defined on a globally hyperbolic Lorentz manifold
could be reduced to the tangent space at a given point, one negleting the gravitational
effects. One finds that the tangent space theory reduces to a free quantum field theory
in a Minkowski space which has local translation invariance and a distinguished invariant state
could be established by a {\em local} unitary mapping. Nevertheless, this unitary operator
depends on the region and there exists no unitary operator which does the mapping for all open
regions simultaneously. Therefore, the problem of how to characterize the physical states
arises. For the discussion of this problem on a general manifold, the setting of the so-called
algebraic approach to quantum field theory (see~\cite{KaWa,Wal,Haag}) is particularly
appropriate, because it treats all states on equal footing, specially that states arising
of unitarily inequivalent representations.

The algebraic approach envolves the theory of $*$-algebras and their states and Hilbert space
representations. In this framework the basic objects are the algebras generated by observables
localized in a given spacetime region. Fields are not mentioned in this setting and are
regarded as a type of coordinates of the algebras. The basic assumption is that {\em all
physical} information must already be encoded in the structure of the local observables. 
Haag and Kastler introduced a mathematical structure for the set of observables of a
physical system by proposing the now so-called Haag-Kastler axioms~\cite{HaKa} for nets of
$C^*$ algebras, later generalized by Dimock~\cite{Dimo} for local observables to
globally hyperbolic manifolds. Recently, a new approach to the model independent description
of quantum field theories has been introduced Brunetti-Fredenhagen-Verch~\cite{BFV},
which incorporates in a local sense the principle of general covariance of general
relativity, thus giving rise to the concept of a locally covariant quantum field theory. The
usual Haag-Kastler-Dimock framework can be regained from this new approach as a special
case.

In this section, we intend to discuss the algebraic formalism so as to 
include supersymmetry on a supermanifold. A straight formulation on a supermanifold can be
performed over the algebraic approach easily, since the construction of the algebra
{\em does not depend ``a priori'' of the manifold}. Let us describe a physical theory in a
general supermanifold from an extended formulation of the ordinary theory in curved spacetime.
An observable algebra can be generated from $\Phi_{\rm sd}(f_{\rm sf})$, 
where $\Phi_{\rm sd}$ are superdistributions (superfields) and $f_{\rm sf}$ test
superfunctions. A complete superalgebra, like above, is represented by
${\mathfrak A_{\rm sa}}={\overline{\bigcup_{{\mathscr O}}
{\mathfrak A}_{\rm sa}({\mathscr O})}}$, where ${\mathfrak A}_{\rm sa}$ denotes the
superalgebra, with ${\mathscr O} \subset {\mathscr M}$ denoting a bounded open
region on a supermanifold $\mathscr M$. We shall assume we have assigned to every bounded
open region $\mathscr O$ in $\mathscr M$ the following properties:

%%%%%%%%%%%%%%%%%%%%%%%%%%%%%%%%%%%%%%%%%%%%%%%%%%%%%%%%%%%%%%%%%%%%%%%%%%%%%%%%%%%%%%%%%
\newcounter{numero}
\setcounter{numero}{0}
\def\Pro{\addtocounter{numero}{1}\item[P.\thenumero]}
\begin{enumerate}

\Pro All ${\mathfrak A}_{\rm sa}({\mathscr O})$
are $*$-superalgebras containing a common unit element, where it is assumed that
the following condition of isotony holds:
\begin{align*}
{\mathscr O}_1 \subset {\mathscr O}_2 \Longrightarrow {\mathfrak A}_{\rm sa}({\mathscr O}_1)
\hookrightarrow {\mathfrak A}_{\rm sa}({\mathscr O}_2)\,\,.
\end{align*}
This condition expresses the fact that the set, which we call in an improper way, of
supersymmetric ``observables'' increases with the size of the localization
region. (Certainly the set of physically interesting observables are obtained
taking the body).  

\Pro  We define the essential notion of locality so that
the restriction of a compact region ${\mathscr O} \in {\mathscr M}$ to a compact region
of the body of the supermanifold, ${\mathscr O}_{\bf b} \in {\mathscr M}_0$, is causally
separated from another compact region ${\mathscr O}^\prime_{\bf b} \in {\mathscr M}_0$.
This implies in the spacelike commutativity,
$[{\mathfrak A}_{\rm sa}({\mathscr O}),{\mathfrak A}_{\rm sa}({\mathscr O}^\prime)]=0$.
We see that this requirement is important, because only with this restriction we can work with
causality: the notion of a suitable proper time curve which intersects the Cauchy surface 
in a global hyperbolic spacetime makes sense only on the body manifold. So, there we can
establish an evolution of Cauchy surfaces to give us a criterion to define a Hadamard form to
the vacuum state. A superdistribution on a supermanifold as a two-point function shows us that
the causality is well-defined in this context. Therefore, we now state: 
if ${\mathscr O}_{\bf b}$ is causally dependent on ${\mathscr O}^\prime_{\bf b}$, 
then ${\mathfrak A}_{\rm sa}({\mathscr O})\subset{\mathfrak A}_{\rm sa}({\mathscr O}^\prime)$.

\Pro  Following Dimock~\cite{Dimo}, we require that there be an
${\mathfrak A}_{\rm sa}({\mathscr O})$ for each supermanifold ${\mathscr M}$ equipped
with some supermetric $g$, which generalizes the Lorentz metric, in a diffeomorphic class.
Let $k:{\mathscr M}_0 \rightarrow {\mathscr M}^\prime_0$ be a $C^\infty$
diffeomorphism on the body manifold, such that $k^*(g_0^\prime)=g_0$, where $g_0$ is
a metric of signature $(+,-,-,-)$ of the body manifold. Then
$z(k):{\mathscr M} \rightarrow {\mathscr M}^\prime$ is a $G^\infty$ superdiffeomorphism 
$z(k)$ from $({\mathscr M},g)$ to $({\mathscr M}^\prime,g^\prime)$ such that
$z(k)^*(g^\prime)=g$, and there is an isomorphism $\alpha_{z(k)}:{\mathfrak A}_{\rm sa}
\rightarrow {\widehat{\mathfrak A}_{\rm sa}}$ such that 
$\alpha_{z(k)}[{\mathfrak A}_{\rm sa}({\mathscr O})]=
{\widehat{\mathfrak A}_{\rm sa}}(z(k)({\mathscr O}))$. One can also show that
$z({\rm id}_{{\mathscr M}_0})={\rm id}_{\mathscr M}$, where
${\rm id}_{{\mathscr M}_0}({\rm id}_{\mathscr M})$ are the identity functions on
${\mathscr M}_0({\mathscr M})$, respectively. Hence, 
$\alpha_{z({\rm id}_{{\mathscr M}_0})}=\alpha_{({\rm id}_{{\mathscr M}})}$ and,
by Eq.(\ref{compo}), we have $\alpha_{z(k_1)}\circ \alpha_{z(k_2)}=\alpha_{z(k_1 \circ k_2)}$. 

\end{enumerate} 
%%%%%%%%%%%%%%%%%%%%%%%%%%%%%%%%%%%%%%%%%%%%%%%%%%%%%%%%%%%%%%%%%%%%%%%%%%%%%%%%%%%%%%%%%%%%%%

It is interesting, in  a particular way, choose a suitable $*$-algebra for a formulation of
quantum fields in connection to the G\r{a}rding-Wightman approach~\cite{Wig}.
In quantum field theory, it is natural to work with tensor product over test functions, since
is usual the presence of more than one field. Therefore, we introduce a tensor algebra of
smooth superfunctions of compact support over ${\mathscr O} \in {\mathscr M}$, where
${\mathscr O}$ is an open region in a supermanifold. Let $f_m$ be a test superfunction
in ${\mathfrak D}_m({\mathscr O})$, so that
$F=\oplus_{m \in {\Bbb N}} f_m(z_1,\ldots,z_m)\in {\mathfrak A}_{\rm sa}({\mathscr O})$,
where here $z_i=(x_i,\theta_i,\bar{\theta}_i)$ denotes the supercoordinates. In a same way we
take $\omega_m(z_1,\ldots,z_m) \in {\mathfrak D}^\prime_m({\mathscr O})$, here
${\mathfrak D}^\prime­_m$ is the dual space of ${\mathfrak D}­_m$ consisting of
$m$-point superdistributions $\omega=\{\omega_m\}_{m \in {\Bbb N}}$,
such that $\omega_m$ belongs to the dual algebra denoted by
${\mathfrak A}^\prime_{\rm sa}({\mathscr O})$.
As we are working on involutive superalgebras, let us define the operation of involution
$(^*)$ by $f^*_m(z_1,\ldots,z_m)=\overline{f_m(z_m,\ldots,z_1)}$,
where $f_m^*=\overline{f_m}$ denotes the complex conjugation.

A superstate $\omega$ in this class of algebra is a normalized positive linear functional
$\omega:{\mathfrak A}_{\rm sa}({\mathscr O})\rightarrow {\mathscr G}_L$, with
$\omega(F^*F)\geq 0$ for all $F \in {\mathfrak A}_{\rm sa}({\mathscr O})$. The normalization
means that $\omega^0=1$. This net of algebra is the Borchers-Uhlmann one~\cite{Bor62}.
Such an algebra does not contain any specific dynamical information, which can be obtained
by specifying a vacuum state on it. Once the vacuum state has been specified, through the
GNS construction which fixes a Hilbert superspace and a vacuum vector, one can extract from
the corresponding time-ordered, advanced or retarded superfunctions the desired information.

A superstate is said to satisfy the essential property of {\em local commutativity}
if and only if for all $m \geq 2$ and all $1 \leq i \leq m-1$ we have
\begin{align*}
\omega_m(f_1 \otimes \cdots \otimes f_i \otimes f_{i+1} \otimes \cdots \otimes f_m)=
\omega_m(f_1 \otimes \cdots \otimes f_{i+1} \otimes f_i \otimes \cdots \otimes f_m)\,\,,
\end{align*}
for all $f_i \in G^\infty_0({\mathscr O})$, such that the restriction of each $f_i$ on compact
regions of the body of supermanifold implies that the
${\rm supp}\,\,f_{i}|_{{\mathscr O}_{\bf b}}$ and
${\rm supp}\,\,f_{i+1}|_{{\mathscr O}_{\bf b}}$ are spacelike separated.
Furthermore, a superstate $\omega$ is ``quasi-free'' if the one-point superdistribution
and all the truncated $m$-point superdistributions for $m \not=2$ vanish, i.e., all
$m$-point superdistributions are obtained from the two-point superdistribution via
relation:
\begin{align*}
&\omega_{2m+1}(f_1 \otimes \cdots \otimes f_m)=0 \quad {\mbox{for $m \geq 0$}}\,\,, \\[3mm]
&\omega_{2m}(f_1 \otimes \cdots \otimes f_m)=
\sum_{\substack{i_1< \cdots <i_{2m} \\
i_k<j_k \\ i_1,\ldots,j_{2m}\,\,{\rm distinct}}} 
\,\,\omega_2(f_{i_1}\otimes f_{j_1}) 
\omega_2(f_{i_2}\otimes f_{j_2}) \cdots \omega_2(f_{i_{2m}}\otimes f_{j_{2m}})\,\,,
\end{align*}
for $m \geq 1$.

It is a well-know result that the physical model can be described by the GNS construction,
showing us how the Hilbert space is constructed and defining what are the operators
(just the algebra representation) acting in this space. According to conventional prescription,
for getting the Hilbert space we choose the quotient between the observable algebra and the
ideal ${\cal N}_{\omega}$ (to guarantee the scalar product existence).
In this stage the problem of several inequivalent representation persists.
In flat superspaces, the super-Poincar\'e invariance of the vacuum state picks out the correct
representation~\cite{Ost}. In general supermanifolds the case is more delicated;
we will look for (super)Hadamard structures. This is motivated by the ordinary general
manifold case. At last, we choose an acceptable Hilbert superspace from the algebraic properties 
via GNS construction by the following identification:
\begin{align*}
\omega_m(f_1 \otimes \cdots \otimes f_m)=
\langle \Omega_\omega, \pi_\omega(f_1) \ldots \pi_\omega(f_m) \Omega_\omega \rangle\,\,, 
\end{align*}
where here $\Omega_\omega$ is a distinguished vector in Hilbert superspace, and
$\pi_\omega$ is the representation of the elements $F \in {\mathfrak A}_{\rm sa}({\mathscr O})$
which play the role of self-adjoint linear operator acting in the Hilbert superspace over test
superfunctions. In addition, we use the physical requirements on the body manifold in order to
define whole set of superstates which are supposed to be distinguished by a certain generalized
form of the spectral condition~\cite{BFK}.

\begin{remark}
The main features of Hilbert superspaces relevant for our purposes are summarized as follows:
$(i)$ when the Grassmann algebra ${\mathscr G}_L$ is endowed with the Rogers norm, every
Hilbert superspace is of the form ${\mathscr H}={\cal H}\otimes {\mathscr G}_L$, where $\cal H$
is an ordinary Hilbert space (the existence of such a subspace ${\cal H}$ of
${\mathscr H}$ called a base Hilbert space is important in physical applications~\cite{Naga}),
$(ii)$ the ${\mathscr G}_L$-valued inner product
$\langle \cdot,\cdot \rangle:{\mathscr H}\times{\mathscr H}\rightarrow {\mathscr G}_L$
respects the body operation $\langle x_{\bf b},y_{\bf b} \rangle=\langle x,y \rangle_{\bf b}$
and $\langle x,x \rangle_{\bf b} \geq 0$ for all $\in {\mathscr H}$, so that
$x \in {\mathscr H}$ has nonvanishing body if and only if
$\langle x,x \rangle_{\bf b} > 0$. For generalizations
of some basic results of the theory of Hilbert space to Hilbert superspaces we refer to the
recent paper~\cite{Rudo} and references therein. $\cqd$
\end{remark}

%%%%%%%%%%%%%%%%%%%%%%%%%%%%%%%%%%%%%%%%%%%%%%%%%%%%%%%
\section{Hadamard (Super)states}
\hspace*{\parindent}
%%%%%%%%%%%%%%%%%%%%%%%%%%%%%%%%%%%%%%%%%%%%%%%%%%%%%%%
As already emphasized, the Hadamard state condition provides a framework in 
which we may improve our understanding to the problem concerning the determination of
physically acceptable states. The motivation for we adopt the Hadamard structure of the
vacuum state in curved spacetime quantum field theory is quite simple. In general, as we
lost the possibility of pick out a good representation for the model due the fact that now
we have not more an invariant structure over the action of an isometry group (in the flat
case, the global Poincar\'e group), we must get another condition of choose. Since we are
able to describe some aspects of a manifold observing the evolution of Cauchy surface (CS)
coming from of asymptotic flat space, a new kind of invariance becomes natural, and this
invariance arises from the preservation of some particular structure while the CS geometry
is changing in determinated manifolds.

In particular, for states whose expectation values of the energy-momentum tensor
operator can be defined by using the point separation prescription for renormalization,
Fulling {\em et al.}~\cite{FuSwWa} showed that if such states have a singularity
structure of the Hadamard form in an open neighbourhood of a Cauchy surface, then they
have their forms preservated independently of the Cauchy evolution. In this case, the
states are said to have the Hadamard form if they can be expressed as
\begin{align*}
\Delta_{\rm Had}(x_1,x_2)=\frac{U(x_1,x_2)}{\sigma(x_1,x_2)} + 
V(x_1,x_2)\,\,{\rm ln} |\sigma(x_1,x_2)|+ W(x_1,x_2)\,\,,
\end{align*}
where $\sigma(x_1,x_2)$ is one-half of the square of the geodesic distance between
$x_1$ to $x_2$. In flat spacetime or in the $x_1 \rightarrow x_2$ limit in curved
spacetime, $\sigma=\frac{1}{2}(x_1-x_2)^2$. It is clear of this that sing supp
$\Delta_{\rm Had}=\{(x_1,x_2)\mid \sigma=\frac{1}{2}(x_1-x_2)^2=0\}$ (we recall that the
singular support of a distribution $u \in {\mathscr D}^\prime(X)$ is the smallest closed
subset $Y$ of $X$ such that $u|_{X \backslash Y}$ is of class $C^\infty$).
$U,V$ and $W$ are regular functions for all choices of $x_1$ and $x_2$.
The functions $U$ and $V$ are geometrical quantities independent of the quantum state,
and only $W$ carries information about the state. Therefore, for free quantum field models in
ordinary globally hyperbolic manifolds, the Hadamard form plays an important role: it is a
strong candidate to describe an acceptable physical representation.

The search for the Hadamard form in the superspace case is simple, since the latter is, in
general, obtainable by applying the function $\delta^2(\bar{\theta}-\bar{\theta^\prime})$
(or $\delta^2({\theta}-{\theta^\prime})$) and an exponential structure
$e^{E(\partial_x,\theta,\bar{\theta})}$
to the ordinary Hadamard form $\Delta_{\rm Had}$ (see Proposition \ref{pro1} below 
and~\cite{WB,CoSc} for details), such that the {\em singularity structure region is not
affected}, i.e., it has a short distance behaviour analogous to the short distance
behaviour discussed in the case of a general spacetime manifold~\cite{PiSi}. This issue
is recaptured in Section \ref{MAS}.
Since we can deal with a supermanifold which has a body manifold being a globally hyperbolic
one (to guarantee this we just report to the construction of Bonora-Pasti-Tonin~\cite{BoPaTo}),
it is important to establish that only {\em projectively} superHadamard structures make sense.
The obvious explanation for this statement is that the structure must cover the global time
notion, and consequently the argument of causality, but over a supermanifold the notion of
causal curves are not well defined unless projectively. The tool to extend the Hadamard
structure to the supersymmetric environment arises from the fact that the existence and
uniqueness of the Grassmannian continuation ($z$-continuation) for $C^{\infty}$ functions is
checked. By a body projection, we always get the ordinary Hadamard structure such that the
latter must be invariant by CS evolution on the body manifold.
This is a consistent result, since we will show in the next section, through an alternative and
equivalent characterization of the Hadamard condition due Radzikowski~\cite{Rad} which involves
the notion of the wavefront set of a superdistribution, that the structure
of singularity is not changed and is condensed in the ordinary region of any Green superfunction,
corroborating to the fact that only on the body of a supermanifold the causality makes sense.

%%%%%%%%%%%%%%%%%%%%%%%%%%%%%%%%%%%%%%%%%%%%%%%%%%%%%%%
\section{Microlocal Analysis in Superspace}
\label{MAS}
\hspace*{\parindent}
%%%%%%%%%%%%%%%%%%%%%%%%%%%%%%%%%%%%%%%%%%%%%%%%%%%%%%%
Important progress in understanding the significance of the Hadamard form relates
it to H\"ormander's concept of wavefront sets and microlocal analysis~\cite{Rad}, 
in a particular way by the wavefront set of their two-point functions. It satisfies
the Hadamard condition if its wavefront set contains only positive frequencies 
propagating forward in time and negative frequencies backward in time.

The focus in this section will be on the extension of the H\"ormander's description 
of the singularity structure (wavefront set) of a distribution to include the
supersymmetric case. The well-known result that the singularities of a superdistribution 
may be expressed in a very simple way through the ordinary distribution 
is proved by functional analytical methods, in particular the methods of 
microlocal analysis formulated in superspace language.

%%%%%%%%%%%%%%%%%%%%%%%%%%%%%%%%%%%%%%%%%%%%%%%%%%%%%%%
\subsection{Standard Facts on Microlocal Analysis}
\hspace*{\parindent}
%%%%%%%%%%%%%%%%%%%%%%%%%%%%%%%%%%%%%%%%%%%%%%%%%%%%%%%
The study of singularities of solutions of differential equations is simplified and the
results are improved by taking what is now known as microlocal analysis. This leads to the
definition of the wavefront set, denoted (${{WF}}$), of a distribution, a refined description
of the singularity spectrum. Similar notion was developed in other versions by 
Sato~\cite{Sa}, Iagolnitzer~\cite{Ia} and Sj\"ostrand~\cite{Sj}. The 
definition, as known nowadays, is due to H\"ormander. He used this 
terminology due to an existing analogy between his studies on the 
``propagation'' of singularities and the classical construction of 
propagating waves by Huyghens. 

The key point of the microlocal analysis is the transference of the study of singularities of
distributions from the configuration space only to the rather phase space, by exploring in
frequency space the decay properties of a distribution at infinity and the smoothness properties
of its Fourier transform. For a distribution $u$ we introduce its wavefront set
${{WF}}(u)$ as a subset in phase space ${\Bbb R}^n\times{\Bbb R}^n$. The
functorially correct definition of phase space is ${\Bbb R}^n\times({\Bbb R}^n)^*$. We shall
here ignore any attempt to distinguish between ${\Bbb R}^n$ and $({\Bbb R}^n)^*$.
We shall be thinking of points $(x,k)$ in phase space as specifying those singular directions
$k$ of a ``bad'' behaviour of the Fourier transform $\widehat{u}$ at infinity that are
responsible for the non-smoothness of $u$ at the point $x$ in position space. So we shall
usually want $k \not= 0$. A relevant point is that ${{WF}}(u)$ is independent of the coordinate
system chosen, and it can be described locally.

As it is well-known~\cite{Hor2,RS2}, a distribution of compact support,
$u \in {\mathscr E}^\prime({\Bbb R}^n)$, is a smooth function if, and only if, its 
Fourier transform, $\widehat{u}$, rapidly decreases at infinity
(i.e., as long as supp$\,u$ does not touch the singularity points). By a fast 
decay at infinity, one must understanding that for all positive integer $N$ 
exists a constant $C_N$, which depends on $N$, such that 
\begin{equation}
|\widehat{u}(k)| \leq (1+|k|)^{-N}C_N\,, 
\qquad \forall\,N \in {\Bbb N};\,\,k \in {\Bbb R}^n\,\,. 
\label{Pepe} 
\end{equation}
If, however, $u \in {\mathscr E}^\prime({\Bbb R}^n)$ is not smooth, then the 
directions along which $\widehat{u}$ does not fall off sufficiently fast 
may be adopted to characterize the singularities of $u$.

For distributions does not necessarily of compact support, still we can verify if its Fourier
transform rapidly decreases in a given region $V$ through the technique of localization.
More precisely, if $V \subset X \subset {\Bbb R}^n$ and $u \in {\mathscr D}^\prime(X)$,
we can restrict $u$ to a distribution $u|_V$ in $V$ by setting $u|_V(\phi)=u(\phi)$, where
$\phi$ is a smooth function with support contained in a region $V$, with $\phi(x)\not=0$,
for all $x \in V$. The distribution $\phi u$ can then be seen as a distribution of compact
support on ${\Bbb R}^n$. Its Fourier transform will be defined as a distribution on ${\Bbb R}^n$,
and must satisfy, in absence of singularities in $V \in {\Bbb R}^n$, the property (\ref{Pepe}).
From this point of view, all development is local in the sense that only the behaviour of the
distribution on the arbitrarily small neighbourhood of the singular point, in the configuration
space, is relevant.
 
Let $u \in {\mathscr D}^\prime({\Bbb R}^n)$ be a distribution and $\phi \in 
C_0^\infty(V)$ a smooth function with support $V \subset {\Bbb R}^n$. Then, 
$\phi u$ has compact support. The Fourier transform of $\phi u$ produces a smooth function in
frequency space. 

\begin{lemma}
 Consider $u \in {\mathscr D}^\prime({\Bbb R}^n)$ and $\phi \in 
C_0^\infty(V)$. Then 
$\widehat{\phi u}(k)=u(\phi e^{-ikx})$.
Moreover, the restriction of $u$ to $V \subset {\Bbb R}^n$ is 
smooth on $V$ if, and only if, for every $\phi \in C_0^\infty(V)$ 
and each positive integer $N$ there exist a constant  
$C(\phi, N)$, which depends on $N$ and $\phi$, such that 
$|\widehat{\phi u}(k)| \leq (1+|k|)^{-N}C(\phi, N)$,
for all $N \in {\Bbb N}$ and $k \in {\Bbb R}^n$. \qed
\label{lema1}
\end{lemma}

If $u \in {\mathscr D}^\prime({\Bbb R}^n)$ is singular in $x$, and $\phi \in 
C_0^\infty(V)$ is $\phi(x)\not= 0$; then $\phi u$ is also singular 
in $x$ and has compact support. However, in some directions in $k$-space 
$\widehat{\phi u}$ until will be asymptotically limited. 
This is called the set of {\em regular directions} of $u$. 

\begin{definition} Let $u(x)$ be an arbitrary distribution,
not necessarily of compact support, on an open set $X \subset {\Bbb R}^{n}$. Then, the set of
pairs composed by singular points $x$ in configuration space and by its associated
nonzero singular directions $k$ in Fourier space 
\begin{align}
{{WF}}(u)=\{(x,k) \in X \times ({\Bbb R}^n\backslash 0)\left|\right. k 
\in \Sigma_x(u)\}\,\,, 
\label{A.2}
\end{align}
is called {\bf wavefront set} of $u$. $\Sigma_x(u)$ is defined to be the 
complement in ${\Bbb R}^n\backslash 0$ of the set of all $k \in {\Bbb 
R}^n\backslash 0$ for which there is an open conic neighbourhood $M$ of $k$ 
such that $\widehat{\phi u}$ rapidly decreases in $M$, for $|k| \rightarrow \infty$. 
\end{definition}

\begin{remarks}
We will now collect some basic properties of the wavefront set:
\begin{enumerate}

\item The ${{WF}}(u)$ is conic in the sense that it remains invariant under the action of 
dilatations, i.e., when we multiply the second variable by a positive 
scalar. This means that if $(x,k) \in {{WF}}(u)$ then
$(x,\lambda k)\in {{WF}}(u)$ for all $\lambda > 0$. 

\item  From the definition of ${{WF}}(u)$, it follows that the projection onto the
first variable, $\pi_1({{WF}}(u)) \rightarrow x$, consists of those points that have no
neighbourhood wherein $u$ is a smooth function, and the projection onto the second variable,
$\pi_2({{WF}}(u)) \rightarrow \Sigma_x(u)$, is the cone around $k$ attached to a such point
de\-no\-ting the set of high-frequency directions responsible for the appearance of a
singularity at this point.

\item The wavefront set of a smooth function is the empty set.

\item For all smooth function $\phi$ with compact suport ${WF}(\phi u)\subset {WF}(u)$.

\item For any partial linear differential operator $P$, with $C^\infty$ coefficients, we have
\[
{WF}(Pu)\subseteq {WF}(u)\,\,.
\]

\item If $u$ and $v$ are two distributions belonging to ${\mathscr D}^\prime({\Bbb R}^n)$,
with wavefront sets ${WF}(u)$ and ${WF}(v)$, respectively; then the wavefront set of
$(u+v) \in {\mathscr D}^\prime({\Bbb R}^n)$ is contained in ${WF}(u)\cup{WF}(v)$.

\item If $U,V$ are open set of ${\Bbb R}^n$, $u \in {\mathscr D}^\prime(V)$, and
$\chi:U \rightarrow V$ a diffeomorphism such that $\chi^*u \in {\mathscr D}^\prime(U)$ is
the distribution pulled back by $\chi$, then ${WF}(\chi^*u)=\chi^*{WF}(u)$. $\cqd$

\end{enumerate}
\end{remarks}

Another result, which we merely state, is needed to complete this briefing on
microlocal analysis.

\begin{theorem}[Wavefront set of pushforwards of a distribution] Let $f:X \rightarrow Y$ be
a submersion, and let $u \in {\mathscr E}^\prime(X)$. Then
\[
WF(f_*u) \subset \{(f(x),\eta) \mid x \in X, (x,^t\!\!f_x^\prime \eta)\in
WF(u)\,\,{\mbox{or}}\,\,^t\!f_x^\prime \eta=0\}\,\,,
\]
where $^t\!f_x^\prime$ denotes the transpose matrix of the Jacobian matrix $f_x^\prime$ of
$f$. \qed
\label{wfspd}
\end{theorem}

%%%%%%%%%%%%%%%%%%%%%%%%%%%%%%%%%%%%%%%%%%%%%%%%%%%%%%%
\subsection{Wavefront set of a Superdistribution}
\hspace*{\parindent}
%%%%%%%%%%%%%%%%%%%%%%%%%%%%%%%%%%%%%%%%%%%%%%%%%%%%%%%
It is already well-known that the singularity structure of Feynman (or more precisely
Wightman) superfunctions is completely associated with the ``bosonic'' sector of the
superspace. Although claims exist that the result is completely obvious, we do not think that
a clear proof is available in the literature, to the best of our knowledge. In fact, there is a
certain gap in the scientific literature between the usual textbook presentation of the
singularity structure of superfunctions and the very mathematical treatement based on
microlocal analysis. The purpose of the present subsection is to fill this gap. As expected,
our result confirms that the decay properties of an ordinary distribution hold also to the case
of a superdistribution, i.e., no new singularity appear by taking into account the structure of
the superspace.

\begin{lemma}
Let $X \subset {\mathscr G}_L^{m,0}$ be an open set, and
$u$ be a superdistribution on $X$ taking values in ${\mathscr G}_L$, i.e., a
linear functional $u:G_0^\infty(X) \rightarrow {\mathscr G}_L$. Let $\phi$ be a
supersmooth function with compact support $K \subset X$. Then $\phi u$ is
also supersmooth on $K$, if its components $(\phi u)(\epsilon(x))$ are smooth on a compact
set $K^\prime \subset \Omega$, where $\Omega$ is the body of superspace. Therefore, the
following estimate holds:
\begin{equation*}
\left| \widehat{\phi u}(k) \right| 
\leq (1+|k_{\bf b}|)^{-N}C(N,\phi)\,\,.
\end{equation*}
\label{main}
\end{lemma}

\begin{proof}[Indication of Proof]
A schematic proof may be constructed along the lines suggested by DeWitt~\cite{DeWitt}: from
Definition \ref{galo} follows that functions of $x$ are in one-to-one correspondence with
functions of $x_{\bf b}$; this implies that in working with integrals over
${\mathscr G}_L^{m,0}$ one may for many purposes proceed as if one were working over
the body of superspace, $\Omega=\{(x,0,0) \in X \mid \epsilon(x) \in {\Bbb R}^m\}$.
Because ${\phi u}(x)$ vanishes at infinity, independently of their souls, the contour
in ${\mathscr G}_{L,0}^m$ may be displaced to coincide with
$\Omega$, without affecting the value of the integral. So, the theory of the Fourier
transforms remains unchanged in form. For the sake of simplicity, we take the case
for which $s(x)=(x-\epsilon(x))$ is a smooth singled-valued function of
$\epsilon(x)=x_{\bf b}$ and $L=2$ is the number of generators of ${\mathscr G}_2^{1,0}$.
This implies 
\begin{align*}
\widehat{\phi u}(k)&=\int dx\,\,e^{ikx}{\phi u}(x) \\[3mm]
&=\int dx_{\bf b}\,\,e^{ik_{\bf b}x_{\bf b}}\left({\phi u}(x_{\bf b})+
i\,x_{\bf b} {\phi u}(x_{\bf b}) k_{ij}\xi^i \xi^j \right)\\[3mm]
&= \widehat{\phi u}(k_{\bf b})+(\widehat{\phi u})^\prime(k_{\bf b})
k_{ij}\xi^i \xi^j\,\,.
\end{align*}
The proof follows one making use of repeated integrations-by-parts generalizing the fact 
$-i\,k_{\bf b}^{-1}\left(\frac{d}{dx_{\bf b}}\,
e^{ik_{\bf b}x_{\bf b}}\right)=e^{ik_{\bf b}x_{\bf b}}$
\begin{align*}
\widehat{\phi u}(k)=\frac{(i)^{|\beta|}}{k_{\bf b}^\beta} 
\left\{\int dx_{\bf b}\,\,e^{-ik_{\bf b}x_{\bf b}}
\left(D_{x_{\bf b}}^{\beta}(\phi u(x_{\bf b}))+
D_{x_{\bf b}}^{\beta}(x_{\bf b}
\phi u(x_{\bf b}))k_{ij}\xi^i \xi^j \right)\right\}\,\,. 
\end{align*}
Taking the absolute value of both sides and using the Banach algebra property of
${\mathscr G}_L$, we get the estimate:
\begin{align}
\left| \widehat{\phi u}(k) \right| 
&\leq \left| \widehat{\phi u}(k_{\bf b}) \right|+
\left| (\widehat{\phi u})^\prime(k_{\bf b}) \right|
\left|k_{ij}\right| \nonumber\\[3mm]
&\leq (1+|k_{\bf b}|)^{-|\beta|}
\left(\sup_{\stackrel{|\beta|\leq m}{x_{\bf b} \in K^\prime}} 
|D_{x_{\bf b}}^{\beta}(\phi u(x_{\bf b}))|+
\sup_{\stackrel{|\beta|\leq m}{x_{\bf b} \in K^\prime}} 
|D_{x_{\bf b}}^{\beta}(x_{\bf b}\phi u(x_{\bf b}))|\left|k_{ij}\right|\right)\,\,.
\label{main1}
\end{align}
This inequality clearly implies our assertion. Hence, in order that (\ref{main1}) be 
smooth, we only need that $\widehat{\phi u}(k)$ be rapidly decreasing as
$|k_{\bf b}| \rightarrow \infty$. The proof may be generalized to include the case in which
$s(x)$ is a multi-valued function of the body and $L$ is finite arbitrarily. We finish the
proof by observing that as expected the soul part of $k$ has a polynomial behaviour.
\end{proof}

\begin{lemma} By replacing ${\mathscr G}_L^{m,0}$ by ${\mathscr G}_L^{m,n}$
in the Lemma \ref{main}, then in this case the following estimate holds:
\begin{align*}
\left| \widehat{\phi u}(k,\theta,\bar{\theta}) \right| 
\leq (1+|k_{\bf b}|)^{-N}C(N,\phi_{(\gamma)})
\|\theta_1\|\|\bar{\theta}_1\|\cdots \|\theta_n\|\|\bar{\theta}_n\|\,\,.
\end{align*}
\label{propo2}
\end{lemma}

\begin{proof}
First, we note that both $u$ and $\phi$ are $G^\infty$ superfunctions which
can be expanded as a polinomial in the odd coordinates whose coefficients are
functions defined over the even coordinates,
\begin{align*}
u(x,\theta,\bar{\theta})=\sum_{(\gamma)=0}^\Gamma z(u_{(\gamma)})(x)
(\theta)^{(\gamma)} \quad {\mbox{and}} \quad 
\phi(x,\theta,\bar{\theta})=\sum_{(\gamma)=0}^\Gamma z(\phi_{(\gamma)})(x)
(\theta)^{(\gamma)}\,\,.
\end{align*}
Then, the proof follows essentially by similar arguments to the proof of the
previous lemma, taking into account the polinomial behaviour of odd variables,
$\theta$ and $\bar{\theta}$. In fact, $\phi u(x,\theta,\bar{\theta})$ is
{\em linear} function in each odd coordinates separately, because each odd
coordinate is nilpotent, and no higher power of a odd coordinate can appear,
i.e., $\phi u(x,\theta,\bar{\theta})$ is an absolutely convergent serie in
the odd coordinates w.r.t. the Rogers norm $\|\cdot\|_1$. Indeed,
$\phi u(x,\theta,\bar{\theta})$ is analytic in the odd coordinates.
This suggests that to take the Fourier transform of
$\phi u(x,\theta,\bar{\theta})$ on the even variables must be sufficient
to infer on the smoothness properties of $\phi u(x,\theta,\bar{\theta})$:
\begin{align}
\widehat{\phi u}(k,\theta,\bar{\theta})&=
\sum_{(\gamma)=0}^\Gamma \sum_{(\mu)=0}^L
(\widehat{\phi u})_{(\gamma),({\mu})}(k_{\bf b})
(\xi)^{(\mu)}(\theta)^{(\gamma)} \nonumber\\[3mm]
&=\sum_{(\gamma)=0}^\Gamma 
\left[\int dx_{\bf b}\,\,e^{ik_{\bf b}x_{\bf b}}
\left(({\phi u})_{(\gamma)}(x_{\bf b})+
i\,x_{\bf b} ({\phi u})_{(\gamma)}
(x_{\bf b}) k_{ij}\xi^i \xi^j+\cdots \right)\right]
(\theta)^{(\gamma)}\,. 
\label{tatu}
\end{align} 
Then, taking the absolute value of both sides of (\ref{tatu}), we obtain from the Banach
algebra property of ${\mathscr G}_L$ and for each integer $N$ the estimate: 
\begin{align}
\left|\widehat{\phi u}(k,\theta,\bar{\theta})\right|=&\left|\sum_{(\gamma)=0}^\Gamma 
\sum_{(\mu)=0}^L(\widehat{\phi u})_{(\gamma),({\mu})}(k_{\bf b})
(\xi)^{(\mu)}(\theta)^{(\gamma)}\right| \nonumber \\
\leq& \sum_{(\gamma)=0}^\Gamma 
\sum_{(\mu)=0}^L\left|(\widehat{\phi u})_{(\gamma),{(\mu})}(k_{\bf b})
\right| \left\|(\theta)^{(\gamma)}\right\| \nonumber \\[3mm]
\leq& (1+|k_{\bf b}|)^{-N}C(N,\phi_{(\gamma)})
\|\theta_1\|\|\bar{\theta}_1\|\cdots \|\theta_n\|\|\bar{\theta}_n\|\,\,.
\end{align}
This proves the lemma.
\end{proof}

So, the odd sector of superspace does not produce any effect on the singular structure of
$u$. Combining the results above, we have proved:

\begin{theorem}
The singularities of  a superdistribution $u$ are located at specific values 
of the body of $x$, the coordinates of the {\bf physical spacetime}, 
independently of the odd coordinates. \qed
\label{prop1}
\end{theorem}

\begin{note}
That the body of the superspace is responsible for carrying all its singular 
structure is not too surprising. Apparently, there exists 
no reason to have superspaces whose topological properties are 
substantially different from its body, which is responsible for carrying all
observables, reflecting some measurable properties of the model. $\cqd$
\end{note}

We sum up the preceding discussion as follows:

\begin{definition}[Wavefront Set of a Superdistribution] The wavefront set $WF(u)$ of a
superdistribution $u$ in a superspace ${\mathscr M}$ is the complement of the set of
all regular directed points in the cotangent bundle $T^*{\mathscr M}_0$,
where ${\mathscr M}_0=\epsilon({\mathscr M})$ is the body of superspace,
excluding the trivial point $k_{\bf b}=0$.
\label{mscs0}
\end{definition}

There is a more precise version of Definition \ref{mscs0}. As we have seen in
Section \ref{DiSmf} all of the foregoing definitions and statements about supermanifolds
may be converted into corresponding definitions and statements about ordinary manifolds,
since associated with a supermanifold ${\mathscr M}$ of dimension $(m,n)$ is a family
of ordinary manifolds, of dimensions $N=2^{L-1}(m+n)$, $(L=1,2,\ldots)$. The resulting
manifold is called the $L$th skeleton of $\mathscr M$ and denoted by
${\cal S}_L({\mathscr M})$~\cite{DeWitt}. With the aid of the family of skeletons we can
define the pushforward (or direct image) of a superdistribution. Let
$X \subset {\cal S}_L({\mathscr M})$ and $Y \subset {\mathscr M}_0$ be open sets and let
$\epsilon$ be the natural projection from ${\cal S}_L({\mathscr M})$ (or ${\mathscr M}$) to
${\mathscr M}_0$, the body map. If we introduce local coordinates $x=(x_1,\ldots,x_N)$ in
$X$, then $Y$ is defined by $x_{\bf b}=(x_1,\ldots,x_m)$. There is a local relationship
between the body and the skeletons given by
\[
{\cal S}_L(X)\overset{\text{diff.}}{=}Y \times {\Bbb R}^{2^{L-1}(m+n)-m}\,\,.
\]
Now, let $u$ be a superdistribution on $X$, then the pushforward $\epsilon_*u$ defined by
$\epsilon_*u(\varphi)=u(\epsilon^*\varphi)$, $\varphi \in C_0^\infty(Y)$, it is a
superdistribution on $Y$. Using these concepts, we can establish the following

\begin{corollary}
Let $\epsilon:X \subset{\cal S}_L({\mathscr M})\rightarrow Y \subset{\mathscr M}_0$ be the
body projection, and let $u \in {\mathfrak D}^\prime(X)$. Then
\begin{align*}
WF(\epsilon_*u)\subset \left\{(x_{\bf b},k_{\bf b}) \in T^*{\mathscr M}_0\backslash 0
\mid\,\exists\,x^\prime=(x_{m+1},\ldots,x_{N^\prime}),
(x_{\bf b},x^\prime,k_{\bf b},0)\in WF(u) \right\}\,,
\end{align*}
where $N^\prime=2^{L-1}(m+n)-m$.
\label{corol}
\end{corollary}

\begin{proof} If $x=(x_{\bf b},x^\prime)$, where $x_{\bf b} \in Y$,
$x^\prime \in {\Bbb R}^{N^\prime}$ and $\epsilon:X \rightarrow Y$ is the body map, then
the Jacobian matrix is of the form $\epsilon^\prime_x=(1,0)$ and the statement follows
by Theorem \ref{wfspd}. Thus, with any superspace ${\mathscr M}$ and body of superspace
${\mathscr M}_0$ the singularities of a superdistribution $\epsilon_*u$ are located in a
natural way in the set of projections of those points of the wavefront set of the
superdistribution $u$ where singular directions are parallel to the $x_{\bf b}$-axis.
\end{proof} 

\begin{example} For the model of Wess-Zumino, which consist of a chiral superfield $\Phi$
in self-interaction, the Feynman superpropagators, in flat superspace, are~\cite{WB}:
\begin{align}
&\Delta^{\rm F}_{\Phi\Phi}(x,\theta,\bar{\theta};
x^\prime,\theta^\prime,\bar{\theta}^\prime)=-i\,m \delta^2(\theta-\theta^{\prime}) 
e^{i(\theta\sigma^\mu\bar{\theta}-\theta^{\prime}\sigma^\mu\bar{\theta}^{\prime})
\partial_\mu}\Delta_{\rm F}(x-x^{\prime})\,\,, \nonumber \es 
&\Delta^{\rm F}_{\bar{\Phi}\Phi}(x,\theta,\bar{\theta};
x^\prime,\theta^\prime,\bar{\theta}^\prime)=
e^{i(\theta\sigma^\mu\bar{\theta}+ 
\theta^{\prime}\sigma^\mu\bar{\theta}^{\prime}-2\theta
\sigma^\mu \bar{\theta}^{\prime})\partial_\mu}\Delta_{\rm F}(x-x^{\prime})\,\,,\label{FS}\es 
&\Delta^{\rm F}_{\bar{\Phi}\bar{\Phi}}(x,\theta,\bar{\theta};
x^\prime,\theta^\prime,\bar{\theta}^\prime)= 
i\,m\delta^2(\bar{\theta}-\bar{\theta}^{\prime}) e^{-i(\theta\sigma^\mu\bar
{\theta}-\theta^{\prime}\sigma^\mu\bar{\theta}^{\prime})\partial_\mu}
\Delta_{\rm F}(x-x^{\prime})\,\,, \nonumber
\end{align}
where $\delta^2(\theta-\theta^{\prime})=(\theta-\theta^{\prime})^2$, with
$x,\theta,\bar{\theta}$ having the form (\ref{snx}) and (\ref{snt}), respectively. According
to our analysis, the wavefront set of Feynman superprogators have the form,
\begin{align*}
WF(\Delta^{\rm F}_{\rm susy})=
\{(x_{\bf b},k_{\bf b};x_{\bf b}^\prime,-k_{\bf b}^\prime;x,0;x^\prime,0)
\mid (x_{\bf b},k_{\bf b};x_{\bf b}^\prime,-k_{\bf b}^\prime) \in 
WF(\Delta^{\rm F}_{\rm susy}|_{{\mathscr M}_0})\}\,\,,
\end{align*}
where ${\rm susy}=(\Phi\Phi;\bar{\Phi}{\Phi};\bar{\Phi}\bar{\Phi})$,
$x=(x_{m+1},\ldots,x_{N^\prime})$, $x^\prime=(x^\prime_{m+1},\ldots,x^\prime_{N^\prime})$,
$\Delta^{\rm F}_{\rm susy}|_{{\mathscr M}_0} \equiv \epsilon_*\Delta^{\rm F}_{\rm susy}$
is the direct image of Feynman superpropagators on the body of superspace, and
$WF(\Delta^{\rm F}_{\rm susy}|_{{\mathscr M}_0})\subset O \cup D$~\cite{Rad},
with the off-diagonal piece given by
\begin{align*}
O=\{(x_{\bf b},k_{\bf b};x_{\bf b}^\prime,-k_{\bf b}^\prime) \in T^*{\mathscr M}_0^2
\mid &(x_{\bf b},k_{\bf b})\sim(x_{\bf b}^\prime,k_{\bf b}^\prime),
x_{\bf b} \not= x_{\bf b}^\prime, \\
&k_{\bf b}\in \overline{V}_{\pm}\quad{\rm if}\quad
x_{\bf b}\in {J}_{\pm}(x_{\bf b}^\prime)\}\,\,,
\end{align*}
where the equivalence relation $(x_{\bf b},k_{\bf b})\sim(x_{\bf b}^\prime,k_{\bf b}^\prime)$
means that there is a lightlike geodesic $\gamma$ connecting $x_{\bf b}$ and
$x_{\bf b}^\prime$, such that at the point $x_{\bf b}$ the covector $k_{\bf b}$ is
tangent to $\gamma$ and $k_{\bf b}^\prime$ is the vector parallel transported along
the curve $\gamma$ at $x_{\bf b}^\prime$ which is again tangent to $\gamma$.

The diagonal piece is given by
\begin{align*}
D=\{(x_{\bf b},k_{\bf b};x_{\bf b},-k_{\bf b}) \in T^*{\mathscr M}_0^2 \backslash 0
\mid x_{\bf b}\in {\mathscr M}_0, k_{\bf b} \in T^*{\mathscr M}_0^2 \backslash 0\}\,\,.
\end{align*}
For this reason, the Feynman superpropagators are singular only for pairs of points
on the body of superspace that can be connected by a lightlike geodesic. $\cqd$
\end{example}

We end this section quoting the main lesson on the microlocal analysis that we can use,
i.e., the one about how the wavefront set may be lifted from superdistributions on open sets
of ${\mathscr G}_L^{m,n}$ to superdistributions on a smooth supermanifold ${\mathscr M}$. 
Such an extension can be achieved in analogy with the ordinary case. Let $\mathscr O$ be
an open neighbourhood of $z \in {\mathscr M}$, which is assumed without loss generality to
be covered by a single coordinate patch, and $u \in {\mathfrak D}^\prime (\mathscr O)$ be
a superdistribution. Then, there exists a diffeomorphism $\chi:{\mathscr O}\rightarrow U
\subset {\mathscr G}_L^{m,n}$, so that $\chi^*u \in {\mathfrak D}^\prime(U)$
is the superdistribution pulled back by $\chi$. Therefore ${WF}(\chi^*u)=\chi^*{WF}(u)$. Now,
let $\phi$ be a supersmooth function with compact
support contained within ${\mathscr O}$ with $\phi(z)\not= 0$ -- one should keep always in mind
that each component $\phi_{(\gamma)}(\epsilon(x))$ of $\phi(z)$ is a smooth function and with
support contained within ${\mathscr O}_{\bf b}$, where ${\mathscr O}_{\bf b}$ denotes an open
neighbourhood of $x_{\bf b} \in {\mathscr M}_0$. Hence, the superdistribution $u\phi$ can be
seen as a superdistribution on ${\mathscr G}_L^{m,n}$ which is of compact support,
and given that there are no points belonging to the ${{WF}}(u)$, the Fourier
transform, $\widehat{u\phi}$, of $u\phi$ is well defined as a superdistribution on
${\mathscr G}_L^{m,n}$ and satisfies the Lemma \ref{propo2}.

%%%%%%%%%%%%%%%%%%%%%%%%%%%%%%%%%%%%%%%%%%%%%%%%%%%%%%%
\section{A Type of Microlocal Spectral Condition}
\hspace*{\parindent}
%%%%%%%%%%%%%%%%%%%%%%%%%%%%%%%%%%%%%%%%%%%%%%%%%%%%%%%
We come back to the question of the Hadamard superstates. As repeatedly stated in this paper,
Hadamard states have acquired a prominent status in connection with the spectral condition,
and are recognized as defining the class of physical states for quantum field
theories on a globally hyperbolic spacetime. Important progress in understanding the
significance of Hadamard states was achieved by Radzikowski (with some gaps filled by
K\"ohler~\cite{Ko}) who succeeded in characterizing the class of these states in terms of the
wavefront set of their two-point function $\omega_2$ satisfying a certain condition. He called
this condition the wavefront set spectral condition (WFSSC). He proposed that a quasifree state
$\omega$ of the Klein-Gordon field over a globally hyperbolic manifold is a Hadamard state if
and only if its two-point distribution $\omega_2$ has wavefront set
\begin{equation}
{WF}(\omega_2)=\left\{(x_1,k_1);(x_2,k_2) \in T^*{\mathscr M}_0^2\setminus \{0\}
\mid (x_1,k_1)\sim(x_2,-k_2)\,\,{\mbox{and}}\,\,k_1^0 \geq 0 \right\}\,\,,
\label{wfssc}
\end{equation}
so that $x_1$ and $x_2$ lie on a single null geodesic $\gamma$, $(k_1)^\mu=g^{\mu\nu}(k_1)_\nu$
is tangent to $\gamma$ and future pointing, and when $k_1$ is parallel transported along
$\gamma$ from $x_1$ to $x_2$ yields $-k_2$. If $x_1=x_2$, we have $k_1^2=0$ and $k_1=k_2$.
Radzikowski in fact showed that this condition is similar to the spectral condition of axiomatic
quantum field theory~\cite{Wig}.

Note that equation (\ref{wfssc}) restricts the singular support of $\omega_2(x_1,x_2)$ to
points $x_1$ and $x_2$ which are null related. Hence, $\omega_2$ must be smooth for all
other points. This is known be true for theory of quantized fields on Minkowski space for
space-like related points. The key is the Bargman-Hall-Wightman theorem which shows that this
obtainable by applying complex Lorentz transformations to the primitive domain of analyticity
determined by the spectral condition. However, a similar prediction on the smoothness does
not exist for time-like related points. Radzikowski suggested to extend the right-hand side
of equation (\ref{wfssc}) to all causally related points, in order to include possible
singularities at time-like related points. 

The microlocal characterization of Hadamard states may be applied equally well to a $n$-point
function, with $n > 2$. This generalization was achieved by Brunetti {\em et al.}~\cite{BFK}.
They suggested a prescription which we recall now. Let ${\cal G}_m$ denotes the set of all
finite graphs~\cite{Grafos}, into some Lorentz manifold ${\mathscr M}_0$,
whose vertices represent points in the set $V=\{x_1,\ldots,x_m\} \in {\mathscr M}_0$, and whose
edges $e$ represent connections between pairs $x_i,x_j$ by smooth curves (geodesics) $\gamma(e)$
from $x_i$ to $x_j$. To each edge $e$ one assigns a covariantly constant causal covector field
$k_e$ which is future directed if $i<j$, but not related to the tangent vector of the curve.
If $e^{-1}$ denotes the edge with opposite direction as $e$, then the corresponding curve
$\gamma(e^{-1})$ is the inverse of $\gamma(e)$, which carries the momentum $k_{e^{-1}}=-k_e$.

\begin{definition}
[$\mu$SC~\cite{BFK}] A state $\omega$ with $m$-point distribution $\omega_m$ is said to
satisfy the Microlocal Spectral Condition if, and only if, for any $m$
\begin{align*}
{{WF}}(\omega_m)\subseteq \Gamma_m\,\,,
\end{align*}
where $\Gamma_m$ is the set $\{(x_1,k_1),\ldots,(x_m,k_m)\}$ for which there exists a graph
$G \in {\cal G}_m$ as described above with $k_i=\sum k_e(x_i)$ where the sum runs over all
edges which have the point $x_i$ as their sources. The trivial momentum configuration
$k_1=\cdots=k_m=0$ is excluded.
\label{msc}
\end{definition}

Passing from a smooth manifold to a smooth supermanifold, it seems reasonable to require
that a superstate satisfies a certain type of microlocal spectrum condition.
A completely analogous statement to the Definition \ref{msc} can be achieved, once more
with the aid of the family of skeletons, ${\cal S}_L(\mathscr M)$, and the graph theory.
Let ${\cal G}_r$ be a set of finite ``{\em supergraphs,}'' into some ${\cal S}_L(\mathscr M)$,
whose vertices represent points in the set $V=\{x_1,\ldots,x_r\} \in {\cal S}_L(\mathscr M)$.
Locally the traditional notion of a supergraph drawing is that its vertices are represented
by points in the hyperplane ${\Bbb R}^{2^{L-1}(m+n)}$, its edges are represented by curves --
that are piecewise linear -- between these points, and different curves meet only in common
endpoints. If $\epsilon_0:{\Bbb R}^{2^{L-1}(m+n)}\rightarrow {\Bbb R}^m$ is the canonical
projection, then $\widetilde{G}=\epsilon_0 G$ is a graphy composed by the projection of those
points of a supergraph whose edges $e$ represent connections between pairs
$x_{{\bf b}_i},x_{{\bf b}_j} \in {\Bbb R}^m$ by curves from $x_{{\bf b}_i}$ to $x_{{\bf b}_j}$.
Then, according to Brunetti {\em et al}~\cite{BFK}, an immersion of a
graph $\widetilde{G}$ into the body manifold ${\mathscr M}_0$ is an assignment of vertices of
$\widetilde{G}$ to points in ${\mathscr M}_0$, and of the edges of $\widetilde{G}$ to piecewise
smooth curves in ${\mathscr M}_0$, $e \rightarrow \gamma(e)$ with source
$s(\gamma(e))=x_{\bf b}(s(e))$ and target $t(\gamma(e))=x_{\bf b}(t(e))$, respectively,
together with a covariantly constant causal covector field $k_{{\bf b}_e}$ on $\gamma$ such
that: ({\em i}) if $e^{-1}$ denotes the edge with opposite direction as $e$, then the
corresponding curve $\gamma(e^{-1})$ is the inverse of $\gamma(e)$; ({\em ii}) for every
edge $e$ the covector $k_{{\bf b}_e}$ is directed toward future if
$x_{\bf b}(s(e)) < x_{\bf b}(t(e))$; ({\em iii}) $k_{{\bf b}_{e^{-1}}}=-k_{{\bf b}_e}$.
Using this construction, we establish:

\begin{definition}[susy$\mu$SC]
A superstate $\omega^{\rm susy}$ with $r$-point superdistribution $\omega_r^{\rm susy}$ is
said to satisfy a Supersymmetric Microlocal Spectral Condition if, and only if, for any $r$
\begin{align*}
{{WF}}(\omega_r^{\rm susy})=\left\{(x_{{\bf b}_1},x^\prime_1,k_{{\bf b}_1},0);\ldots;
(x_{{\bf b}_r},x_r^\prime,k_{{\bf b}_r},0) \mid {{WF}}(\epsilon_*\omega_r^{\rm susy})
\subseteq \widetilde{\Gamma}_r \right\}\,\,,
\end{align*}
where $\widetilde{\Gamma}_r$ is the set $\{(x_{{\bf b}_1},k_{{\bf b}_1});\ldots;
(x_{{\bf b}_r},k_{{\bf b}_r})\}$ for which there exists a graph
$\widetilde{G}$ as described above with $k_{{\bf b}_i}=\sum k_{{\bf b}_e}(x_{{\bf b}_i})$ 
where the sum runs over all edges which have the point $x_{{\bf b}_i}$ as their sources.
The trivial momentum configuration $k_{{\bf b}_1}=\cdots=k_{{\bf b}_r}=0$ is excluded.
\label{mscs}
\end{definition}

\begin{remarks} We would like to call attention to two important points:
\begin{itemize}

\item The Definition \ref{mscs} indicates that for a superstate $\omega^{\rm susy}$
the (susy$\mu$SC) is e\-qui\-va\-lent to the requirement that all of the component fields
satisfy the microlocal spectral conditions~\cite{BFK} on the body manifold. This observation
is significant because it is in agreement with the DeWitt's remark which asserts that,
in physical applications of supersymmetric quantum field theories, the spectral condition of
the GNS-Hilbert superspace is restricted to the ordinary GNS-Hilbert space that sits inside
the GNS-Hilbert superspace.

\item The Definition \ref{mscs} provides us with a ``global'' microlocal spectral condition.
In our setting the word ``global'' means that the singular support of all component fields
is embodied in ${{WF}}(\epsilon_*\omega_m^{\rm susy})$. This is typical feature of
supersymmetric theories in superspace language. For instance, for the chiral superfield of
Wess-Zumino~\cite{WB}, in analogy to the scalar component field, the Hadamard condition for
a spinorial component field is formulated in terms of its two-point distribution $\omega_2$.
The latter are obtainable by applying the adjoint of the spinorial operator to a suitable
auxiliary Hadamard state of the squared spinorial equation. For fixed spinor indices the
wavefront set of the latter is contained in r.h.s. of equation (\ref{wfssc}) and derivatives
do not enlarge the wavefront set. $\cqd$

\end{itemize}

\end{remarks}

Next we give a example of an application of our definiton. We restrict ourselves to the
simplest case of massive chiral/antichiral fields of the Wess-Zumino model in flat
superspace, leaving other cases as the Wess-Zumino model, or supersymmetric gauge theories
in curved superspace for future works.

%%%%%%%%%%%%%%%%%%%%%%%%%%%%%%%%%%%%%%%%%%%%%%%%%%%%%%%%%%%%%%%%%%%
\subsubsection*{$\star$ The Free Wess-Zumino Model in Flat Superspace}
\hspace*{\parindent}
%%%%%%%%%%%%%%%%%%%%%%%%%%%%%%%%%%%%%%%%%%%%%%%%%%%%%%%%%%%%%%%%%%%
The simplest $N=1$ supersymmetric model in four dimension is the free model of
Wess-Zumino~\cite{WB}, which consists of a chiral superfield 
$\Phi(x,\theta,\bar{\theta})$, resp. antichiral superfield 
$\bar{\Phi}(x,\theta,\bar{\theta})$, obeying the differential constraint 
$\bar{D}_{\dot{\alpha}} \Phi=0$, resp. ${D}_\alpha \bar{\Phi}=0$. As usual, 
\begin{equation}
D_\alpha=\frac{\partial}{\partial \theta^\alpha}-i\sigma^\mu_{\alpha 
\dot{\alpha}}\bar{\theta}^{\dot{\alpha}}\partial_\mu\,,\quad
\bar{D}_{\dot{\alpha}}=-\frac{\partial}{\partial \bar{\theta}^{\dot{\alpha}}}+
i\theta^\alpha\sigma^\mu_{\alpha \dot{\alpha}}\partial_\mu\,\,, 
\label{Sec1.eq4}
\end{equation}
is a supersymmetric covariant derivatives. Our notations and conventions 
are those of~\cite{PiSi}. The elements of the $N=1$ superspace are parametrized by
even and odd coordinates $z^M=(x^\mu,\theta^\alpha,\bar{\theta}^{\dot{\alpha}})$, with 
$\mu=(0,\ldots,3),\,\alpha=(1,2),\,\dot{\alpha}=(\dot{1},\dot{2})$, 
where $\theta$ and its complex conjugate $\bar{\theta}$, are odd 
coordinates and by construction they anticommute with each other. In this case the body
manifold is ${\Bbb R}^m$ and the body map is the augmentation map
$\epsilon:{\mathscr G}_L^{m,n} \rightarrow {\Bbb R}^m$.

The superfield $\Phi(z)$ is a function mapping superspace into the even part of a
Grassmann algebra~\cite{Rogers}. With the help of the commutation rule
$\bar{D}_\alpha\,\,(e^{-i \theta \sigma^\mu \bar\theta \partial_\mu}\phi)=
e^{-i \theta \sigma^\mu \bar\theta \partial_\mu}(-{\partial}/{\partial
\bar{\theta}^\alpha})\phi$, the chiral superfield can be expanded in powers of the
odd coordinates as 
\begin{equation}
\Phi(z)= e^{-i \theta \sigma^\mu 
\bar\theta \partial_\mu}(\varphi(x)+\theta \psi(x)+ \theta^2 F(x))\,\,, 
\label{Sec1.eq3}
\end{equation}
with $\varphi \overset{\text{def}}{=}2^{-1/2}(A+iB)$ and $F \overset{\text{def}}{=}
2^{-1/2}(D-iE)$. $A$, $B$ and $\psi$ are respectively the scalar, pseudoscalar and
spin-$1/2$ physical component fields of $\Phi$, whereas $D$ and $E$ are their scalar 
and  pseudoscalar auxiliary components. The latter are necessary for a classical off-shell
closure of the supersymmetry algebra (they do not corresponding to propagating degrees of
freedom in that appear through non-derivative terms).

As above, the antichiral superfield $\bar{\Phi}(z)$, with the help of the 
commutation rule ${D}_\alpha\,\,(e^{i\theta \sigma^\mu \bar\theta\partial_\mu}\phi)=
e^{i \theta \sigma^\mu \bar\theta \partial_\mu}({\partial}/{\partial {\theta}^\alpha})\phi$,
can be expanded in component fields:
\begin{equation}
\bar{\Phi}(z)= e^{i\theta \sigma^\mu \bar\theta\partial_\mu}(\varphi^*(x)+
\bar{\theta} \bar{\psi}(x)+ \bar{\theta}^2 F^*(x))\,\,. 
\label{Sec1.eq3a}
\end{equation}

The quantum version of the Wess-Zumino model is based on the classical field equations 
\begin{equation}
\frac{1}{16}\bar{D}^2 \bar{\Phi} + \frac{m}{4}\Phi=0\,,\quad 
\frac{1}{16}{D}^2 \Phi + \frac{m}{4}\bar{\Phi}=0\,\,. 
\label{Sec1.eq1}
\end{equation}                                      
Applying the operator $D^2$ to the first equation (resp. $\bar{D}^2$ to the 
second equation), multiplying the second equation by $4m$ (resp. the first 
equation), and using the commutation relation 
$[D^2,\bar{D}^2]=8iD\sigma^\mu\bar{D}\partial_\mu+16\Box$; one may combine 
them in order to find 
\begin{equation}
(\Box_x + m^2)\Phi=0\,,\quad (\Box_x + m^2)\bar{\Phi}=0\,\,. 
\label{Sec1.eq2}
\end{equation}

To our classical superfields $\Phi$ and $\bar{\Phi}$, we associate quantum superfields, an
operator-valued ``superdistributions,'' smeared with ``supertest'' functions, 
\begin{align}
F(z)&= e^{-i \theta \sigma^\mu 
\bar\theta \partial_\mu}(f(x)+\theta \chi(x)+ \theta^2 
h(x))\,\,,\nonumber\es
\bar{F}(z)&= e^{i \theta \sigma^\mu 
\bar\theta \partial_\mu}(f^*(x)+\bar{\theta} \bar{\chi}(x)+ 
\bar{\theta}^2 h^*(x))\,\,, 
\label{Sec1.eq5}
\end{align}
with $F(z), \bar{F}(z) \in G_0^\infty(U,{\mathscr G}_L)$, the ${\mathscr G}_L$-valued
superfunctions on an open set $U \subset {\mathscr G}_L^{m,n}$ which have compact
support. 

For all $F(z),G(z)\in G_0^\infty(U,{\mathscr G}_L)$, we define the commutation relations 
\begin{align}
\left[\Phi(\bar{F}),\Phi(\bar{G})\right]=&
\int d\mu(z) d\mu(z^\prime)\,\,
\Delta^{\rm PJ}_{\Phi\Phi}(z,z^\prime)\bar{F}(z)\bar{G}(z^\prime)
\,\,,\nonumber\\
\left[\bar{\Phi}(F),\Phi(\bar{G})\right]=&
\int d\mu(z) d\mu(z^\prime)\,\,
\Delta^{\rm PJ}_{\bar{\Phi}\Phi}(z,z^\prime)F(z)\bar{G}(z^\prime)
\,\,,\\
\left[\bar{\Phi}(F),\bar{\Phi}(G)\right]=&
\int d\mu(z) d\mu(z^\prime)\,\,
\Delta^{\rm PJ}_{\bar{\Phi}\bar{\Phi}}(z,z^\prime)F(z)G(z^\prime)
\,\,.\nonumber
\label{Sec1.eq6}
\end{align}
where $d\mu(z) \overset{\text{def}}{=}d^8z=d^4x d^2\theta d^2\bar{\theta}$. We call 
$\Delta^{\rm PJ}_{\Phi\Phi}$, $\Delta^{\rm PJ}_{\bar\Phi\Phi}$ and 
$\Delta^{\rm PJ}_{\bar\Phi \bar\Phi}$ the Pauli-Jordan superdistributions, 
fundamental solutions of the homogeneous equations (\ref{Sec1.eq2}). In 
fact they are two-point distributions, elements of ${\mathfrak D}^\prime(U)$. 

The vacuum expectation value of the product $\Phi(F)\Phi(G)$ satisfies the 
relation 
\begin{align}
\left(\Omega,\Phi(F)\Phi(G)\Omega\right)=
\left(w_2^{\rm susy}(z,z^\prime),\,F(z) 
G(z^\prime)\right)\,\,. 
\end{align}
The distribution $w_2^{\rm susy}(z,z^\prime)$ extends the Wightman 
formalism. For this reason, we call $w_2^{\rm susy}(z,z^\prime)$ Wightman 
superdistribution of two-points. 

The Wightman superdistribution of $n$-points will be symbolically written 
under the form~\cite{Ost}: 
\begin{equation}
w_n^{\rm susy}\left(z_1,,\ldots,z_n\right) =\left(\Omega,\Phi 
\left(x_1;\theta_1,\bar{\theta}_1\right) \ldots 
\Phi \left(x_n;\theta_n,\bar{\theta}_n\right) \Omega\right)
\,\,, \label{wightman1}
\end{equation}
and
\begin{equation}
w_n^{\rm susy}\left(F_n\right)=\int \prod_{i=1}^nd\mu_i\,\,w_n^{\rm 
susy}\left(z_1,\ldots,z_n\right) F_n\left(z_1,\ldots,z_n\right)\,\,. 
\label{wightman2} 
\end{equation}
In this definition, we have fixed the order in which we take the superdistribution
and the supertest function. 

\begin{proposition} -- The two-point Hadamard, Pauli-Jordan and Wightman superdistributions
have the following dependence in $x,\theta,\bar{\theta}$: 
\begin{align}
&\Delta^{X}_{\Phi\Phi}(x,\theta,\bar{\theta};
x^\prime,\theta^\prime,\bar{\theta}^\prime)= -i\,m \delta^2(\theta-\theta^{\prime}) 
e^{i(\theta\sigma^\mu\bar{\theta}-\theta^{\prime}\sigma^\mu\bar{\theta}^{\prime})
\partial_\mu}\Delta_{X}(x-x^{\prime})\,\,,\nonumber\es 
&\Delta^{X}_{\bar{\Phi}\Phi}(x,\theta,\bar{\theta};
x^\prime,\theta^\prime,\bar{\theta}^\prime)= e^{i(\theta\sigma^\mu\bar{\theta}+ 
\theta^{\prime}\sigma^\mu\bar{\theta}^{\prime}-2\theta
\sigma^\mu \bar{\theta}^{\prime})\partial_\mu}\Delta_{X}(x-x^{\prime})\,\,,\es 
&\Delta^{X}_{\bar{\Phi}\bar{\Phi}}(x,\theta,\bar{\theta};
x^\prime,\theta^\prime,\bar{\theta}^\prime)= 
i\,m\delta^2(\bar{\theta}-\bar{\theta}^{\prime}) e^{-i(\theta\sigma^\mu\bar
{\theta}-\theta^{\prime}\sigma^\mu\bar{\theta}^{\prime})\partial_\mu}
\Delta_{X}(x-x^{\prime})
\,\,,\nonumber 
\end{align}
where $X=(\rm Had,\rm PJ,\rm W)$. 
\label{pro1}
\end{proposition}

\begin{proof}[Idea of Proof.] We start from (\ref{FS}) and use the fact that
in terms of even and odd solutions of the homogeneous wave equation, the
function $\Delta_{\rm F}(x-x^{\prime})$ can be write as
\begin{equation}
\Delta_{\rm F}(x-x^{\prime})=\frac{1}{2}\left[i\,\Delta_{\rm Had}(x-x^{\prime})
+ \varepsilon(x^0-x^{0\prime})\Delta_{\rm PJ}(x-x^{\prime})\right]
\label{FHPJ}
\end{equation}
Then, by replacing (\ref{FHPJ}) in (\ref{FS}), we immediately get the Hadamard and
Pauli-Jordan superdistribution as stated. The Wightman superdistribution is obtained
directly from the fact the $\Delta_{\rm PJ}(x-x^{\prime})=\Delta_{\rm W}(x-x^{\prime})-
\Delta_{\rm W}(x-x^{\prime})$ and $\Delta_{\rm Had}(x-x^{\prime})=-i
(\Delta_{\rm W}(x-x^{\prime})+\Delta_{\rm W}(x-x^{\prime}))$. 
\end{proof}

\begin{proposition} Let $\omega^{\rm susy}$ be a state for the quantum Wess-Zumino model
on flat superspace, whose $r$-point superdistributions $\omega^{\rm susy}_r$ satisfy the
Wightman axi\-oms~\cite{axiomas}. Then $\omega^{\rm susy}$ satisfies
the Definiton \ref{mscs}.
\label{scwz}
\end{proposition}

\begin{proof}
This is an immediate consequence of Corollary \ref{corol} above and Theorem 4.6 of~\cite{BFK}.
\end{proof}

%%%%%%%%%%%%%%%%%%%%%%%%%%%%%%%%%%%%%%%%%%%%%%%%%%%%%%%
\section{Final Considerations}
\hspace*{\parindent}
%%%%%%%%%%%%%%%%%%%%%%%%%%%%%%%%%%%%%%%%%%%%%%%%%%%%%%%
Having proposed an extension of some structural aspects that have successfully been
applied in the development of the theory of quantum fields 
propagating on a general spacetime manifold so as to include superfield models
on a supermanifold, it would be interesting to consider the perturbative treatment of
interacting quantum superfield models, in particular the formulation
of renormalization theory on supermanifolds. The main problem which still remains in this
rather restrictive framework is the mathematically consistent definition of all 
powers of Wick ``superpolynomials'' and their time-ordered products for the 
noninteracting theory, which serve as building blocks for a perturbative 
definition of interacting superfields. Another work devoted to its solution is in
progress~\cite{DanCaio}, such that covariance with respect to supersymmetry is
manifestly preserved. The renormalization scheme underlying our construction is the one of
Epstein-Glaser. It is formulated, unlike the other renormalization schemes, in configuration
space. Therefore, it becomes appropriate to define carefully perturbative renormalization on a
generic spacetime manifold. Recently, Brunetti and Fredenhagen~\cite{BF} (with some gaps filled
by Hollands and Wald~\cite{HW}) have shown that the Wick polynomials and their time-ordered
products can be defined in globally hyperbolic spacetimes. By the methods
of this paper we can define powers of Wick ``superpolynomials'' and their time-ordered products
for the noninteracting theory.

%%%%%%%%%%%%%%%%%%%%%%%%%%%%%%%%%%%%%%%%%%%%%%%%%%%%%%%%%%%%%%%%%%%%
\section*{Acknowledgements}
\hspace*{\parindent}
%%%%%%%%%%%%%%%%%%%%%%%%%%%%%%%%%%%%%%%%%%%%%%%%%%%%%%%%%%%%%%%%%%%%%
We would like to thank Profs. O. Piguet and J.A. Helay\"el-Neto
for some helpful comments on a preliminary draft of this paper. 
C.M.M. Polito is supported by the Brazilian agency CAPES.

%%%%%%%%%%%%%%%%%%%%%%%%%%%%%%%%%%%%%%%%%%%%%%%%%%%%%%%%%%%%%%%%%%%%%%%%%%
%\newpage
%%%%%%%%%%%%%%%%%%%%%%%%%%%%%%%%%%%%%%%%%%%%%%%%%%%%%%%%%%%%%%%%%%%%%%%%%%


\begin{thebibliography}{99}

\bibitem{Carlip} S. Carlip, ``Quantum Gravity: a Progress Report,'' gr-qc/0108040.

\bibitem{Haw} S.W. Hawking, Commun. Math. Phys. {\bf 43} 199 (1975).

\bibitem{Wald} R.M. Wald, Commun. Math. Phys. {\bf 54} 1 (1977).

\bibitem{WiBr} B.S. DeWitt and R.W. Brehme, Ann. Phys. {\bf 9} 220 (1960).

\bibitem{Ful} S.A. Fulling, {\it Aspects of Quantum Field Theory in 
Curved Space-Time} (Cambridge University Press, 1989).

\bibitem{KaWa} B.S. Kay and R.M. Wald, Phys. Rep. {\bf 207} 49 (1991). 

\bibitem{Wal} R.M. Wald, {\it Quantum Field Theory in Curved Spacetime and
Black Hole Thermodynamics} (The University of Chicago Press, 1994).

\bibitem{Rad} M.J. Radzikowski, Commun. Math. Phys. {\bf 179} 529 (1996); 
Commun. Math. Phys. {\bf 180} 1 (1996).

\bibitem{Kay} B.S. Kay, ``Quantum Field Theory on Curved Space-Time,'' In
{\it Differential Geome\-trical Methods in Theoretical Physics,} edited by 
K. Bleuler and M. Werner, (Dordrecht: Kluwer Academic Publishers, 1988), pp. 
373--393.

\bibitem{Ko} M. K\"ohler, ``The Stress Energy Tensor of a
Locally Supersymmetric Quantum Field on a Curved Spacetime,'' Doctoral 
dissertation, University of Hamburg, 1995, gr-qc/9505014;
Class. Quantum Grav. {\bf 12} 1413 (1995).

\bibitem{BFK} R. Brunetti, K. Fredenhagen and M. K\"ohler,
Commun. Math. Phys. {\bf 180} 663 (1996).
 
\bibitem{Hor1} L. H\"ormander, Acta Math. {\bf 127} 79 (1971).

\bibitem{DH} J.J. Duistermaat and L. H\"ormander, Acta Math. {\bf 128} 183 (1972).

\bibitem{Jun} W. Junker, ``Adiabatic Vacua and Hadamard States
for Scalar Quantum Fields on Curved Spacetime,'' Doctoral dissertation, 
University of Hamburg, 1995, hep-th/9507097.  

\bibitem{KRW} B.S. Kay, M.J. Radzikowski and R.M. Wald,
Commun. Math. Phys. {\bf 183} 533 (1997). 

\bibitem{BF}  R. Brunetti and K. Fredenhagen, ``Interacting
Quantum Fields in Curved Space: Renormalizability of $\varphi^4$,''
gr-qc/9701048; ``Interacting Quantum Fields on a Curved Background,'' 
hep-th/9709011; Commun. Math. Phys. {\bf 208} 623 (2000).

\bibitem{Holl} S. Hollands, Commun. Math. Phys. {\bf 216} 635 (2001).

\bibitem{Kra} K. Kratzert, Annalen Phys. {\bf 9} 475 (2000). 

\bibitem{Ver} R. Verch, Commun. Math. Phys. {\bf 205} 337 (1999);
``On Generalizations of the Spectrum Condition,'' math-ph/0011026.

\bibitem{SaVe} H. Sahlmann and R. Verch, Rev. Math. Phys. {\bf 13} 1203 (2001).

\bibitem{Wi} E. Witten, J. Diff. Geom. {\bf 17} 661 (1982).

\bibitem{Jaf} A. Jaffe, A. Lesniewski and K. Osterwalder,\
Commun. Math. Phys. {\bf 178} 313 (1987).

\bibitem{SeWi} N. Seiberg and E. Witten, Nucl. Phys. B {\bf 426} 19 (1994).

\bibitem{Se} N. Seiberg, Nucl. Phys. B {\bf 435} 129 (1995).

\bibitem{Rogers} A. Rogers, J. Math. Phys. {\bf 21} 1352 (1980).

\bibitem{CaReTe} R. Catenacci, C. Reina and P. Teofilatto,
J. Math. Phys. {\bf 26} 671 (1985).

\bibitem{Bry} P. Bryant, Math. Proc. Camb. Phil. Soc. {\bf 107} 501 (1990).

\bibitem{BoPaTo} L. Bonora, P. Pasti and M. Tonin, J. Math. Phys. {\bf 23} 839 (1982).

\bibitem{NK} S. Nagamachi and Y. Kobayashi, Lett. Math. Phys. {\bf 15} 17 (1988).

\bibitem{HaKa} R. Haag and D. Kastler, J. Math. Phys. {\bf 5} 848 (1964).

\bibitem{Dimo} J. Dimock, Commun. Math. Phys. {\bf 77} 219 (1980);
Trans. Amer. Math. Soc. {\bf 269} 133 (1982).

\bibitem{DeWitt} B.S. DeWitt, {\it Supermanifolds}
(Cambridge Monographs on Mathematical Physics, Second Edition, 1992).

\bibitem{JaPi} A. Jadczyk and K. Pilch, Commun. Math. Phys. {\bf 78} 373 (1981).

\bibitem{HoQuRaUr} J. Hoyos, M. Quir\'os, J. Ram\'\i rez Mittelbrunn
and F.J. de Urr\'\i res, J. Math. Phys. {\bf 25} 833, 841 and 847 (1984).

\bibitem{BrCi} U. Bruzzo and R. Cianci, Commun. Math. Phys. {\bf 95} 393 (1984).

\bibitem{RaCr} J.M. Rabin and L. Crane, Commun. Math. Phys. {\bf 100} 141 (1985). 

\bibitem{Rabin} J.M. Rabin, ``Supermanifolds and Super Riemann Surfaces,''
{\it Lectures given at the NATO Advanced Research Worhshop 
on Super Field Theory, Vancouver, 1986}.

\bibitem{BrCi1} U. Bruzzo and R. Cianci, Lett. Math. Phys. {\bf 11} 21 (1986).

\bibitem{Bruzzo} U. Bruzzo, ``Field Theories on Supermanifolds: General Formalism,
Local Supersymmetry, and the Limit of Global Supersymmetry,''
in {\it Proceedings, Topological Properties and Global Structure of Space-time,
Erice, 1985}, pp.21-29.

\bibitem{Vlavolo} V.S. Vladimirov and I.V. Volovich,
Theor. Math. Phys. {\bf 59} 317 (1984).

\bibitem{Rudo} O. Rudolph, Commun. Math. Phys. {\bf 214} 449 (2000).

\bibitem{Hor2} L. H\"ormander, {\it The Analysis of Linear Partial 
Differential Operators I} (Springer Verlag, Second Edition, 1990).

\bibitem{Folhas} The leaves of a 
soul foliation accumulate if there exist two points $s_+, s_-$ and a sequence $F^{(n)}$
such that for arbitrary transverse submanifolds $\Sigma_+$ through $s_+$ and $\Sigma_-$
through $s_-$, there is some $F^{(n_o)}$, with $F^{(n_o)} \cup \Sigma_+ \not=\emptyset$ 
and $F^{(n_o)} \cup \Sigma_- \not= \emptyset$.

\bibitem{Geroch} R. Geroch, J. Math. Phys. {\bf 9} 1739 (1968);
J. Math. Phys. {\bf 11} 343 (1970).

\bibitem{BrFree} P. Breitenlohner and D.Z. Freedman, Ann. Phys. {\bf 144} 249 (1982).

\bibitem{Treves1} F. Treves, {\it Topological Vector Spaces, Distributions
and Kernels} (Academic Press, 1967).

\bibitem{Haag} R. Haag, {\it Local Quantum Physics: Fields
Particles, Algebras} (Springer, 1996, Second Edition).

\bibitem{BFV} R. Brunetti, K. Fredenhagen and R. Verch,
Commun. Math. Phys. {\bf 237} 31 (2003).

\bibitem{Wig} R.F. Streater and A.S. Wightman, {\it PCT, Spin-Statistics and
All That} (Benjamin, New York, 1964).

\bibitem{Bor62} H.J. Borchers, Nuovo Cimento {\bf 24} 214 (1962).

\bibitem{Ost} K. Osterwalder, ``Supersymmetric Quantum Field Theory,'' Workshop on
{\it Constructive Results in Field Theory and Statistical Mechanics and Condensed
Matter Physics, Palaiseau, France, July 1994} (Springer LNP 446).

\bibitem{Naga} S. Nagamachi, J. Math. Phys. {\bf 33} 4274 (1992).

\bibitem{FuSwWa} S.A. Fulling, M. Sweeny and R.M. Wald,
Commun. Math. Phys. {\bf 63} 257 (1978).

\bibitem{WB} J. Wess and J. Bagger, {\it Supersymmetry and Supergravity}
(Second Edition, Princeton University Press, 1992).

\bibitem{CoSc} F. Constantinescu and G. Scharf, ``Causal Approach to Supersymmetry:
Chiral Superfields,'' hep-th/0106090.

\bibitem{PiSi} A comprehensive account on the divergences of superpropagators and 
the renormalization of supersymmetric theories can be found in the textbook of O. Piguet
and K. Sibold, {\it Renormalized Supersymmetry: The Perturbative Theory of $N=1$
Supersymmetry Theories in Flat Space-Time} (Progress in Physics, vol 12, Birkh\"auser, 1986).

\bibitem{Sa} M. Sato, ``Hyperfunctions and Partial Differential 
Equations,'' {\it Conf. on Funct. Anal. and Related Topics, Tokyo, 1969,} pp31-40.

\bibitem{Ia} D. Iagolnitzer, ``Microlocal Essential Support of a 
Distribution and Decomposition Theorems -- An Introduction,'' in 
{\it Hyperfunctions and Theoretical Physics} (Springer LNM 449, 1975). 

\bibitem{Sj} J. Sj\"ostrand, ``Singularit\'es Analytiques 
Microlocales,'' {\it Ast\'erisque, 1982}.

\bibitem{RS2} M. Reed and B. Simon, {\it Fourier Analysis,
Self-Adjointness}'' (Academic Press, 1975).

\bibitem{Grafos} A graph is a pair $G=(V,E)$, where the elements of $V$ are the
{\em vertices} (or {\em nodes}, or {\em points}) of the graph $G$, and the elements of $E$ are
its {\em edges} (or {\em lines}). The numbers of vertices of a graph is its order, and graphs
are finite or infinite according to their order.

\bibitem{axiomas} Some work concerning the axiomatic supersymmetric quantum field theory
is contained in~\cite{Ost,Consta}, where it is shown that the standard Wightman axioms of
a relativistic quantum field theory can be modified so as to allow supersymmetry.

\bibitem{Consta} F. Constantinescu, Lett. Math. Phys. {\bf 62} 111 (2002).

\bibitem{DanCaio} D.H.T. Franco and C.M.M. Polito, 
``Renormalization of the Wess-Zumino Model on a BPT-Supermanifold,'' work in progress.

\bibitem{HW} S. Hollands and R. Wald, Commun. Math. Phys. {\bf 223} 289 (2001);
Commun. Math. Phys. {\bf 231} 309 (2002).

\end{thebibliography}
\end{document}